\documentclass[prb, twocolumn, superscriptaddress, floatfix, tighten]{revtex4-2}

\usepackage{xcolor}
\usepackage{amsmath}
\usepackage{amssymb}
\usepackage{bm}      	
\usepackage{graphicx}
\usepackage[unicode=true,pdfusetitle,
 bookmarks=false,bookmarksnumbered=false,bookmarksopen=false,
 breaklinks=false,pdfborder={0 0 1},backref=false,colorlinks=true,urlcolor=blue,citecolor=blue,linkcolor=black]
 {hyperref}
\usepackage{subfigure}
\usepackage{textcomp}
\usepackage{microtype} 
\usepackage{braket}

\newcommand{\bsub}{\begin{subequations}}
\newcommand{\esub}{\end{subequations}}

\newcommand{\vex}[1]{\bm{\mathrm{#1}}}
\newcommand{\e}{\varepsilon}

\numberwithin{equation}{section}

\numberwithin{equation}{section}

\begin{document}

\title{Critical Filaments and Superconductivity in Quasiperiodic Twisted Bilayer Graphene} 
\def\rice{Department of Physics and Astronomy, Rice University, Houston, Texas
77005, USA}
\def\rcqm{Rice Center for Quantum Materials, Rice University, Houston, Texas
77005, USA}
\author{Xinghai Zhang}\affiliation{\rice}
\author{Justin H. Wilson}
\affiliation{Department of Physics and Astronomy, Louisiana State University, Baton Rouge, LA 70803, USA}
\affiliation{Center for Computation and Technology, Louisiana State University, Baton Rouge, LA 70803, USA}
\author{Matthew S. Foster}\affiliation{\rice}\affiliation{\rcqm}
\date{\today}

\begin{abstract}
Multilayer
moir\'e materials can exhibit 
topological
electronic  
features yet are inherently quasiperiodic---leading to wave function interference whose 
Anderson-localizing
tendency can be mitigated by topology.
We consider a quasiperiodic variant of the chiral Bistritzer-MacDonald model for twisted bilayer graphene with two incommensurate moir\'e potentials that serves as a toy model for twisted trilayer. We observe ``filaments'' linking magic angles with enhanced density of states and fractal wave functions that evade localization; states away from the filaments mimic fractal surface states of dirty topological superconductors. We demonstrate that topological quasiperiodicity can broadly enhance superconductivity \emph{without} magic-angle fine-tuning.
\end{abstract}

\maketitle

\section{Introduction}

Amplifying electronic interactions and uncovering topological phases drives much of the research in moir\'e materials \cite{MacDonald19}.
Correlated phases including superconductivity were observed in twisted bilayer graphene (TBLG) \cite{MacDonald19}, attributed to enhanced interactions due to flattening of the moir\'e bands near magic angles \cite{Santos07,Li2009,BM11,Tarnopolsky19}.
Concurrently, topology
is 
enabled in TBLG by the coupling of same-chirality Dirac electrons between the layers \cite{Po18,Po19,Song19,Ahn19}.

Generic twist angles in moir\'e materials
lead to incommensurate (quasiperiodic) tunneling,
but this has negligible effects in TBLG.
By contrast,
quasiperiodicity is ubiquitous when \emph{multiple} moir\'e patterns compete, as occurs in 
systems 
such as twisted trilayer graphene (TTLG),
where superconductivity was recently observed
\cite{Uri23,Substrate}.
Previous work on 
quasiperiodic
Dirac and moir\'e materials 
focused on tight-binding models
with lattice incommensuration
\cite{Pixley2018, Chou2020, Fu2020},
and provided glimpses of a kind of wave-function criticality linked with ``magic angles.''
Wave-function criticality (multifractality) is
typically associated to \emph{disorder-driven} Anderson localization transitions \cite{Evers08}.
Quasiperiodicity induces interference that can also drive Anderson localization, 
although the most-studied cases in one dimension (1D) show features different from random 1D systems
\cite{AA80,Sokoloff85}.
At the same time
the tendency towards localization can be mitigated by topology,
which protects boundary and bulk states in nontrivial bands.

The \emph{interplay} between multifractality 
and interactions can 
produce surprising effects.
For example, the average Cooper pairing amplitude can be \emph{enhanced} near a superconductor-insulator transition if 
Coulomb interactions are externally screened.
The mechanism involves the Russian-doll-like nesting (``Chalker scaling'') of 
multifractal states,
and can occur
near the localization transition~\cite{Feigelman07,Feigelman10} 
and in 1D quasiperiodic models \cite{Tezuka2010,Fan2021,Zhang22}.

\begin{figure}[t!]
    \centering
    \includegraphics[width=0.8\columnwidth]{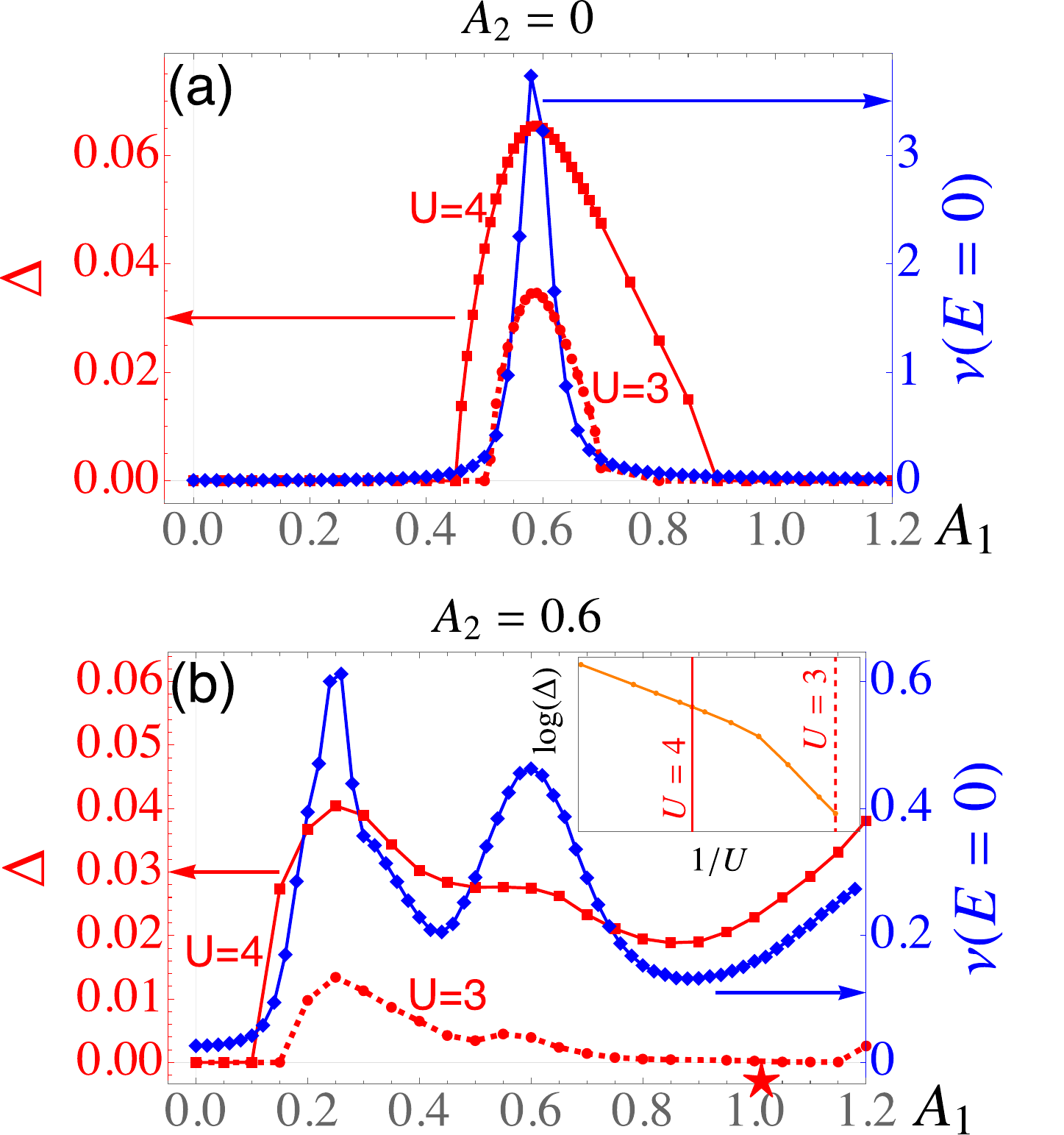}
    \caption{(a): Pairing amplitude $\Delta$ and density of states (DOS) $\nu(E)$ for the chiral Bistritzer-MacDonald (BM) model; the enhancement of superconductivity occurs near the magic-angle 
    ($A_1\approx 0.586$).
    (b): The same for the quasiperiodic BM model; superconductivity is broadly enhanced by wave-function fractality \emph{without} magic-angle fine-tuning.
    The parameters $A_{1,2}$ are the strengths of the two incommensurate BM potentials, 
    see Eqs.~(\ref{eq:hqBM}) and (\ref{eq:Vq}).
    All results are obtained from self-consistent BCS numerics using the kernel polynomial method (Sec.~\ref{sec:KPM}); $U$ is the interaction strength measured relative to the kinetic energy.
    The inset shows the $U$-dependence of $\Delta$ at $A_1=1$ and $A_2=0.6$. 
    Here the combination of Chalker scaling and topology \cite{Ghorashi18,Karcher21,UFO24} leads to \emph{BCS-like} dependence  
    $\Delta \sim e^{-c/U}$ 
    (see Secs.~\ref{sec:MRSC} and \ref{sec:PoEE}),
    instead of the threshold behavior of a 2D Dirac semimetal. The latter occurs away from the magic angle in 
    (a).}
    \label{fig:SC_qBM_BM}
\end{figure}

This raises a question: Can
quasiperiodic and topological
effects in 
multilayer
moir\'e materials lead to criticality that enhances correlated behavior, such as superconductivity?
We address this by studying a quasiperiodic variant of the chiral Bistritizer-MacDonald (BM) model for TBLG.
Our model incorporates two incommensurate BM potentials as occurs in TTLG, and serves as a toy version of the latter with two layers instead of three.
As we show in this paper, quasiperiodicity with topology can lead to criticality and amplify
pairing \emph{without} the need to fine-tune to magic angles,
see Fig.~\ref{fig:SC_qBM_BM}. 

The BM model for 
TBLG
\cite{Santos07,BM11} possesses an emergent particle-hole symmetry \cite{Song19}, falling into class C in the $10$-fold classification of topological matter \cite{Ryu10} and 2D Dirac Hamiltonians \cite{BernardLeClair02}.
If AA tunneling vanishes, a chiral symmetry emerges \cite{Tarnopolsky19}.
This chiral BM model takes the same form as the \emph{2D surface theory} of a bulk class CI topological superconductor (TSC) \cite{Schnyder08}, while the BM model corresponds to the same surface, subject to orbital time-reversal symmetry breaking.

Quenched disorder, which respects the chiral symmetry, leads to remarkable results for class-CI 
TSC
surface states, including a power-law vanishing density of states (DOS) 
$\nu(E) \sim \left|E\right|^{1/7}$ \cite{Nersesyan94} and 
critical 
low-energy wave functions with universal multifractal statistics \cite{CIMFC}. Recent results have demonstrated that a more radical kind of universal wave-function criticality can extend over a wide range of the energy spectrum for these 
surfaces
(``spectrum-wide quantum criticality,'' SWQC \cite{Ghorashi18,Karcher21,UFO24}).

In this paper, we show that topological quasiperiodicity leads to DOS enhancements and wave-function fractality.
We establish that quasiperiodic multilayer graphene effectively simulates \emph{disordered} TSC surface states.
The resulting 
SWQC
and absence of Anderson localization are exotic 
TSC features
\cite{Ghorashi18, Karcher21}
that are
synthetically realizable in 
multilayer
moir\'e materials.
Our results are surprising because they precisely link the criticality of dirty surface states of bulk topological phases to quasiperiodic layered 2D materials. Since the bulk is ``missing,'' one might say that quasiperiodic twisted multilayers \emph{holographically} realize bulk topological superconductor classes.

We 
study the consequences of topologically-protected quasiperiodicity for superconductivity using self-consistent BCS calculations \cite{Zhang22}, finding 
enhanced 
Cooper
pairing \cite{Feigelman07,Feigelman10} without fine-tuning to magic angles
(Fig.~\ref{fig:SC_qBM_BM}). 
For simplicity, we consider 
intravalley pairing. 
Our
calculations incorporate additional aspects beyond those of the normal state such as Altshuler-Aronov corrections, which can also cause Anderson localization; 
such corrections shift the Bogoliubov-de Gennes eigenstate localization transition in a 1D quasiperiodic model \cite{Zhang22}. 
We demonstrate that this effect is absent here due to topological protection \cite{Xie2015}, and we find that the superfluid stiffness precisely encodes the topological character of the normal state
(another instance of the ``holography''). 
Our approach can be applied to the more complicated case of TTLG \cite{Zhang25}.

\subsection{Outline}

This paper is organized as follows. 
The main results are presented and discussed in Sec.~\ref{sec:MainResults}; we first introduce our quasiperiodic version of the chiral Bistritzer-MacDonald model and detail the setup for the numerical kernel-polynomial method (KPM) calculations. We then summarize results for the density of states and wave-function multifractality, linking the latter to TSC surface states. Finally, we consider superconductivity and discuss further directions. 
The rest of the paper contains technical details.
Sec.~\ref{sec:KPM} provides a pedagogical introduction to the KPM method, particularly the implementations needed for wave function calculations, self-consistent BCS theory, and superfluid stiffness computation. 
Sec.~\ref{sec:MoreDetail} gives additional results and details for the density of states and wave-function criticality.
Sec.~\ref{sec:SilverRatio} summarizes results for a silver-ratio variant quasiperiodic model.
Finally, Sec.~\ref{sec:SCDetails} collects additional details and results for superconductivity in the quasiperiodic chiral Bistritzer-MacDonald model.


\section{Main results \label{sec:MainResults}}

\subsection{Model}

We consider a quasiperiodic 
version of the chiral
Bistritzer-MacDonald (qBM) model \cite{BM11,Tarnopolsky19},
\begin{equation}
    h(\mathbf{r})
    = 
    -iv_F \, \boldsymbol{\sigma}\cdot \nabla 
    +
    V_{\mathbf{q}} (\mathbf{r}) 
    +
    V_{\mathbf{q}'}(\mathbf{r}) \,.
    \label{eq:hqBM}
\end{equation}
The Pauli matrices $\sigma_{1,2}$
act on 
the sublattice space, and $V_{\mathbf{q}}(\mathbf{r})$ and $V_{\mathbf{q}'}(\mathbf{r})$ 
denote moir\'{e} potentials 
coupling the two layers in 
AB- and BA-stacking regimes,
\begin{equation}
    V_{\mathbf{q}}(\mathbf{r}) 
    = A_1 
    \left[ 
    U_0 (\mathbf{r}) \, \tau^+ \sigma^+ 
    + 
    U_1(\mathbf{r}) \, \tau^+ \sigma^- 
    + 
    \textrm{H.c.}
    \right]\,.
    \label{eq:Vq}
\end{equation}
Here 
$\tau^\pm = (\tau_x \pm i \tau_y)/2$ 
(and similarly for $\sigma^\pm$) with 
$\tau_j$ as the Pauli matrices that link the layers.
The potentials $U_0$ and $U_1$ take the forms \cite{BM11}
$
    U_0(\mathbf{r}) 
    = 
    e^{-i\mathbf{q}_1\cdot \mathbf{r}}
    +
    e^{-i\mathbf{q}_2\cdot \mathbf{r}}
    +
    e^{-i\mathbf{q}_3\cdot \mathbf{r}}
$
and 
$
     U_1(\mathbf{r}) 
    = 
    e^{-i\mathbf{q}_1\cdot \mathbf{r}}
    +
    e^{-i\frac{2\pi}{3}}
    e^{-i\mathbf{q}_2\cdot \mathbf{r}}
    +
    e^{i\frac{2\pi}{3}}
    e^{-i\mathbf{q}_3\cdot \mathbf{r}}\,.
$
Here 
$\mathbf{q}_1=k_\theta (1, 0)$ and 
$\mathbf{q}_{2/3} = k_\theta (-1/2, \pm \sqrt{3}/2)$
are the reciprocal lattice vectors of the 
usual moir\'e potential.
The momentum $k_\theta = 2K \sin\theta/2$ defines the moir\'e scale 
(with $K$ the K-point wavevector modulus in graphene).
The second moir\'{e} potential $V_{\mathbf{q}'}(\mathbf{r})$ takes the same form but with different reciprocal vectors 
$\{\mathbf{q}'_i\}$ 
and potential strength $A_2$.
When the two sets 
$\{\mathbf{q}_i\}$ and $\{\mathbf{q}'_i\}$ 
are mutually incommensurate,
we have a quasiperiodic model. 
In this 
paper,
we focus on the case $\mathbf{q}_i = \beta \, \mathbf{q}_i'$ with 
$\beta$
the inverse golden ratio $(\sqrt{5}-1)/2$; we have verified that similar results obtain for other (e.g., silver) ratios, see Sec.~\ref{sec:SilverRatio}.

\begin{figure}[t!]
    \centering
    \includegraphics[width=\columnwidth]{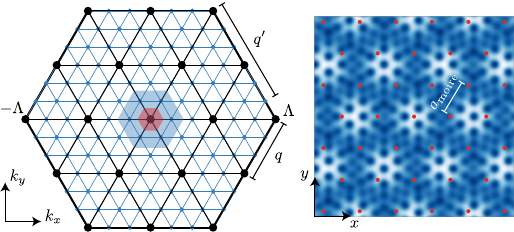}
    \caption{(left) An illustration of the $k$-space grid used for $q'/q = 5/3$ and $N_\Lambda  = 4$. The blue region is the moir\'e Brillouin zone produced by $A_1$ and the red region is the super-moir\'e Brillouin zone of the combined potentials for this commensurate fraction. (right) A density plot of $\lvert\lvert V_{\mathbf q}(\mathbf r) + V_{\mathbf q'}(\mathbf r)\rvert\rvert^2$ with $A_1 = A_2$ to illustrate the quasiperiodicity. The red dots are the original moir\'e sites for the $A_1$ potential, separated by the scale 
    $a_{\text{moir\'e}} = {4\pi}/{\sqrt 3 k_\theta}$.}
    \label{fig:illustration2D}
\end{figure}

The qBM model is realized in momentum space using a momentum cutoff $\Lambda = N_\Lambda k_\theta/2$. 
Focusing on $V_\mathbf q(\mathbf r)$, the moir\'e Brillouin zone (mBZ) has linear size $k_\theta$, and we keep $2N_\Lambda^2$ moir\'e bands.
Without a second potential, a momentum space lattice of linear size $N_\Lambda$ is sufficient;
however, $\mathbf q'_i$ allows hoppings \emph{within} the mBZ. 
Since $|\mathbf q_i'| = k_\theta / \beta$, we controllably reach larger system sizes with rational approximations of $\beta$, i.e.\ $\beta_n = F_n/F_{n+1}$ for Fibonacci numbers $F_n$. 
In this case, the denominator $\frac{|\mathbf q_i'|}{k_\theta} = \frac{F_{n+1}}{F_n}$ determines the linear size of momentum space points required within one mBZ.
Therefore, the full momentum space lattice has linear size $N_\Lambda F_n$ and spacing $\Delta k = k_\theta/F_n$;
in the incommensurate limit ($n\rightarrow\infty$), all momentum can be accessed.
In the following, we work in units such that
$v_F=1$ and $k_\theta = \left|\mathbf{q}_i\right|=1$;
all energies are measured in units of $v_F \, k_\theta$.
Our setup is 
illustrated in 
Fig.~\ref{fig:illustration2D}.


\subsection{Density of states and fractal wave functions}

\begin{figure}[t!]
  \centering
  \includegraphics[width=\columnwidth]{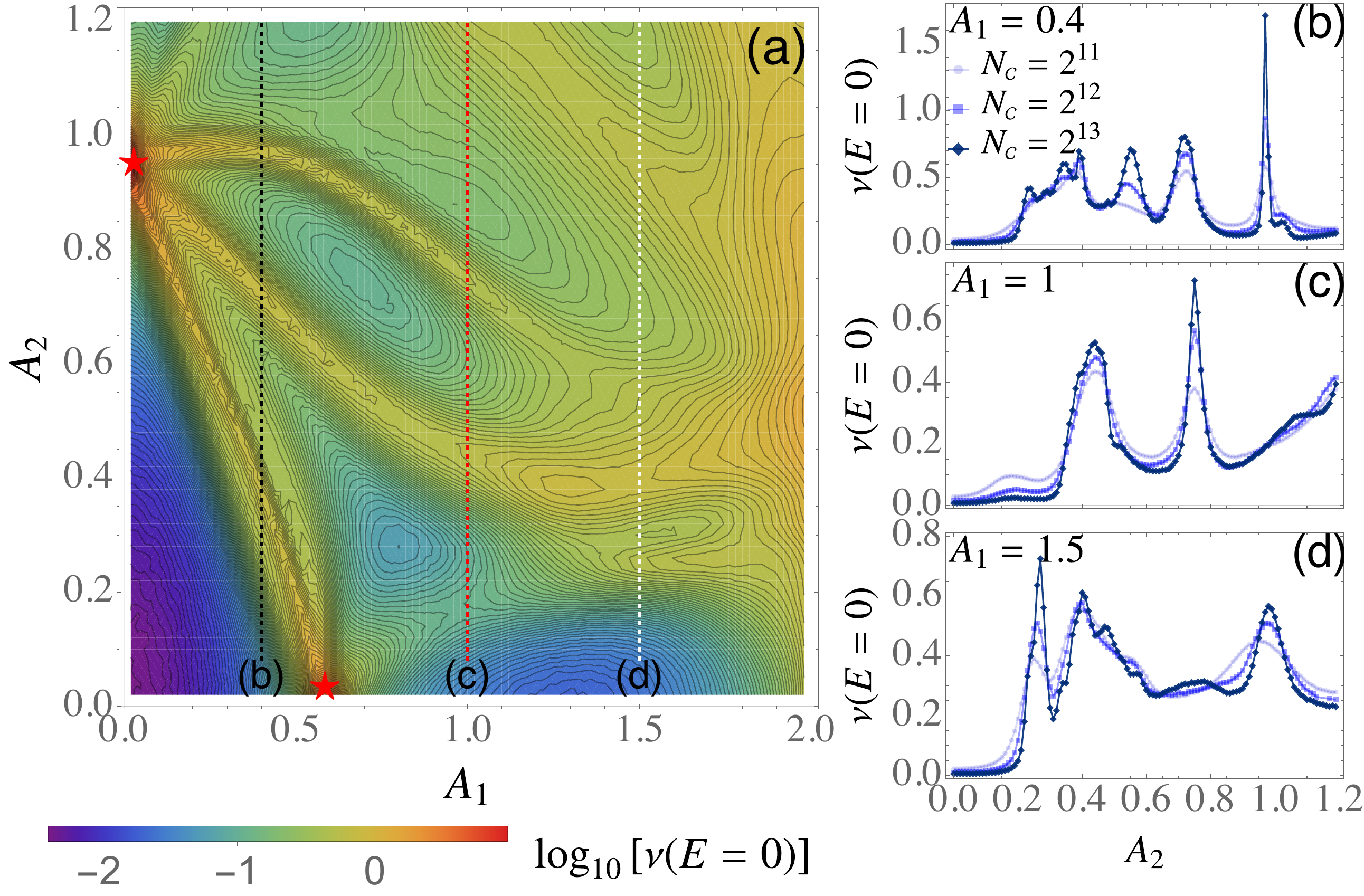}
\caption{The zero-energy density of states $\nu(0)$ 
for the qBM model in chiral limit. Here $A_1$ and $A_2$ are the strengths of the two sets of moir\'{e} potentials. (a): The map of $\nu(0)$ for the two moir\'e potentials. The zero-energy DOS has large peaks at the magic angles of the $A_1$ and $A_2$ potentials [indicated by red stars in (a)]. The magic angles of $A_1$ and $A_2$ are connected by several filaments, where $\nu(0)$ peaks.  
(b), (c) and (d): $\nu(0)$ along the cuts at $A_1=0.4$, $1$ and $1.5$, respectively. The peaks correspond to the filaments and $\nu(0)$ remains nonzero between these. This is our first hint that quasi-periodicity has driven a single-particle phase transition away from the semimetal in these regions. The data are obtained with KPM for a momentum lattice with  $N_\Lambda = 17$ and $F_n=89$. We use Chebyshev expansion order $N_{\scriptscriptstyle \mathcal{C}}=2^{11}$ for the DOS map (a) and $N_{\scriptscriptstyle \mathcal{C}}=2^{11}$, $2^{12}$ and $2^{13}$ in (b-d).}
  \label{fig:dos-map}
\end{figure}

Dirac cones have a vanishing DOS $\nu(E) \sim | E|/v^2$ with velocity $v$; at the magic angle $v \rightarrow 0$ and $\nu(0)$ diverges.
In the quasiperiodic case, at small $A_{1,2}$ the semimetal is stable and eigenstates remain plane-wave-like \cite{VM2020}. 
However, when $\nu(0) > 0$ (or vanishes with a small power-law), 
multifractality--typically associated to an Anderson metal-insulator transition~\cite{Evers08}--can occur.
To reveal these (potential) single-particle phases, Fig.~\ref{fig:dos-map} shows $\nu(0)$ for the qBM model evaluated with the kernel polynomial method (KPM)~\cite{KPM2006}.
Fig.~\ref{fig:dos-map}(a) shows the $\nu(0)$ landscape while $A_{1,2}$ vary. 
There are large lakes with $\nu(0)>0$ separated by \emph{filaments} where the DOS is strongly enhanced.
The filaments connect the magic angles (red stars) of the monochromatic potentials; both filaments and lakes host multifractal wave functions (see below). 
Fig.~\ref{fig:dos-map}(b-d) show particular cuts.
In the regime close to the $A_1$ or $A_2$ axes away from the magic angles, the system remains semimetallic and $\nu(0)\rightarrow 0$.  
By contrast, $\nu(0)$ peaks at the filaments and either remains nonzero or vanishes with a nontrivial power-law between the filaments, see Sec.~\ref{sec:MoreDetail}.
In the regions of comparable $A_{1}$ and $A_{2}$, between the filaments, $\nu(0)$ is generally nonzero; in fact, as we show below, the system exhibits SWQC behavior in these regions associated to class-CI surface states \cite{Ghorashi18}.


Both cutoff $\Lambda$ and number of moir\'e unit cells $F_n$ determine the momentum space lattice, 
which need to both be large; therefore, to determine wave functions, we employ the Chebyshev filtering and shaking method \cite{Pieper2016, Guan2021}.
This method obtains eigenfunctions in a small energy window, which could contain thousands of states near 
a narrow
band, see Sec.~\ref{sec:KPMWF}.
We use the method to obtain eigenfunctions for the qBM model with $\sim 10^6$ lattice sites.

\begin{figure}[b!]
  \centering
  \includegraphics[width=0.48\textwidth]{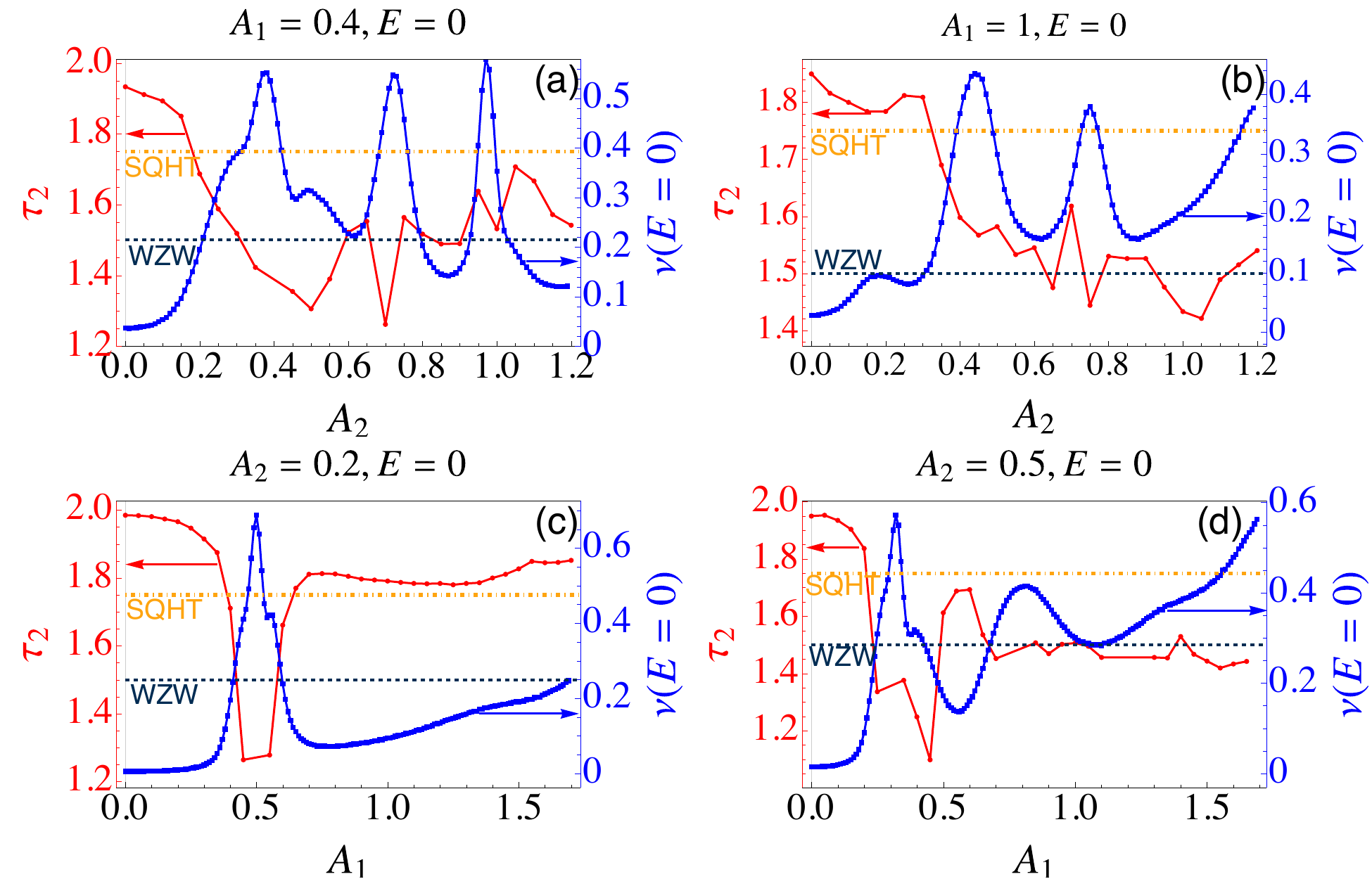}
  \caption{The second multifractal dimension $\tau_2$ averaged for states near zero energy and zero energy DOS $\nu(E=0)$ for the qBM model. (a) and (b): along the vertical cuts at $A_1=0.4$ and $1$, respectively; (c) and (d): along the horizontal cuts at $A_2=0.2$ and $A_2=0.5$, respectively. The navy blue dashed and orange dot-dashed lines indicate $\tau_2=3/2$ for the WZW states and $\tau_2=7/4$ for the SQHT states, respectively. }
  \label{fig:tau2-A2}
\end{figure}

\begin{figure*}
  \centering
  \includegraphics[width=0.98\textwidth]{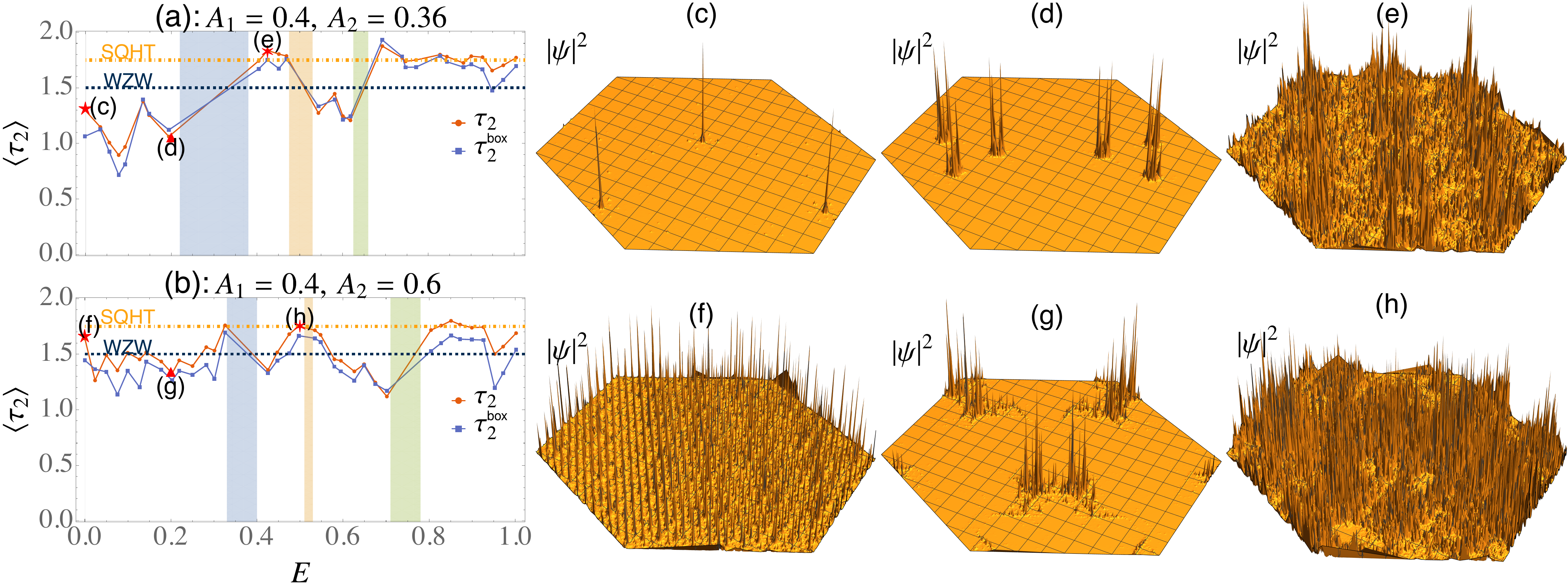}
  \caption{The second multifractal dimension $\tau_2$ versus energy and the wave functions at the one of filaments and in the lakes between the filaments. (a): $\tau_2$-$E$ at the filament with $A_1=0.4$ and $A_2=0.36$; (b): same as (a) but in the lakes between filaments with $A_1=0.4$ and $A_2=0.6$.
  (c)-(e): the wave functions at the filament for states at $E=0$, $0.2$ and $0.425$; 
  (f)-(h): the wave functions in the lake for states at $E=0$, $0.2$ and $0.5$. 
  The shadows in (a) and (b) show the energy gaps in the spectrum. The navy dashed and orange dot-dashed line in (a) and (b) show the reference value of $\tau_2$ for the WZW and SQHT states, respectively.
  Here we $N_\Lambda=11$, $F_n=55$ and the wave functions reside on a $605\times 605$ lattice in real space.}
  \label{fig:tau2-wf}
\end{figure*}

The statistics of single-particle wave functions can be characterized by the inverse participation ratio (IPR)
$
    P_q 
    \equiv
    \sum_i
    \left|\psi_i\right|^{2q}\,.
$
Here $\psi_i$ is the (normalized) wave function in real space, obtained by Fourier transform of the momentum-space wave function.
For extended wave functions, $P_q\sim L^{-d(q-1)}$, with $L$ 
the linear dimension of the system; for localized wave functions, $P_q \sim L^0$.
Further, if $P_q \sim L^{-\tau_q}$ and $0<\tau_q<d(q-1)$, that wave function is \emph{critical}.
In the thermodynamic limit ($L\to \infty$), the multifractal dimension can be evaluated via
$
    \tau_q = -\log{P_q}/\log{L}\,.
$
The $\tau_q$ can also be obtained with the box-averaged IPR,
$
    P_q^{\scriptscriptstyle \mathsf{box}}
    = \sum_{i_b} 
    \left(\sum_{\text{box}}
    \left|\psi_i\right|^2\right)^q
    \propto 
    b^{\tau_q}\,,
    \label{eq:Pqbox}
$
with $b$ 
the linear size of the box. 
We use both methods to calculate the $\tau_q$ and they are consistent 
for the appropriate 
box sizes. 

The second multifractal dimension $\tau_2$ 
for low-energy wave functions
is shown in Fig.~\ref{fig:tau2-A2} for the qBM model along the cuts $A_1=0.4$ and $1$. 
Corresponding $\tau_3$ plots appear in Sec.~\ref{sec:MoreDetail}.
The value
$\tau_2 \approx 2$ indicates ballistic wavefunctions, and 
Fig.~\ref{fig:tau2-A2}(a) shows 
such
states 
for small $A_2$ and $A_1=0.4$. 
We observe
a correlation between 
$\nu(0)$ and multifractality;
at the first filament (with increasing $A_2$), $\tau_2$ decreases, fluctuates for stronger $A_2$, and 
dips near the 
additional
filaments.
The value of $\tau_2$ near the filaments 
in Fig.~\ref{fig:tau2-A2}(a,c)
indicates that the wave functions 
there
are strongly multifractal, 
see also Fig.~\ref{fig:tau2-wf}(c,d,g).
Note that Fig.~\ref{fig:tau2-A2}(c) follows the pattern for magic-angle semimetals as outlined in \cite{Chou2020, Fu2020}.
In other regimes, such as between the filaments, the wave functions are also multifractal with a different value of 
$\tau_2$. 
In the nonperturbative regime, e.g., Fig.~\ref{fig:tau2-A2}(b) along $A_1=1$, $\tau_2$ 
fluctuates
around $3/2$ and is only partially correlated with $\nu(0)$.
However, at this point $\tau_2$ appears to have saturated to the known value $\tau_2=3/2$ for 
zero-energy 
Wess-Zumino-Witten (WZW)
surface states of a disordered class-CI TSC
\cite{CIMFC,Evers08,BernardLeClair02}. 
[Fig.~\ref{fig:tau2-A2}(d) also shows 
this saturation.]

Finite-energy surface-state wave functions of a dirty TSC can also 
evade localization via SWQC, with critical statistics tied to a quantum-Hall plateau transition \cite{Ghorashi18,Sbierski20,Karcher21,UFO24}.
For the class-CI qBM model, we anticipate finite-energy states with SWQC associated 
to the \emph{spin} quantum-Hall transition (SQHT) \cite{Ghorashi18,Karcher21}.
To compare with finite energies in the qBM model, we take two separate values, on the filament ($A_1=0.4$ and $A_2=0.36$) and between the filaments ($A_1=0.4$ and $A_2=0.6$) [see Fig.~\ref{fig:tau2-wf}(a,b)], and examine a range of energies, extracting $\tau_2$.
On the filament [Fig.~\ref{fig:tau2-wf}(a)], the low-energy states are strongly multifractal (with $\tau_2\approx 1$) with a few isolated peaks [energies $E=0,0.2$, 
Figs.~\ref{fig:tau2-wf}(c,d)]; the higher energies oscillate between SQHT states [$\tau_2\approx 7/4$ and less strongly peaked, see Fig.~\ref{fig:tau2-wf}(e)] and strongly multifractal states. 
The filaments, originating from the flat bands at the magic-angles, have hybridized those states strongly into these multifractal states. 
In between the filaments [Fig.~\ref{fig:tau2-wf}(b)], the low-energy 
wave functions
are less multifractal than on the filament with WZW-like states [$\tau_2\approx 3/2$ and $\tau_3\approx 5/2$, Fig.~\ref{fig:tau2-wf}(g)] and the high-energy states that
reside
deep within moir\'e bands are SQHT states [$\tau_2\approx 7/4$ and $\tau_3\approx 13/4$, Fig.~\ref{fig:tau2-wf}(h)].
Near the energy gaps, the wave functions 
show stronger fractal features.


\subsection{Superconductivity \label{sec:MRSC}}

With multifractality established and linked to the enhanced density of states, we are motivated by results that Cooper pairing amplitudes can be enhanced \cite{Feigelman07,Feigelman10,Tezuka2010,Fan2021,Zhang22} to look at superconductivity in these regimes.

The Bogoliubov-de Gennes (BdG)
Hamiltonian for the qBM model with pseudospin, layer, and spin-singlet
$s$-wave pairing 
is given by
$
    H = 
    \int d\mathbf{r} 
    \,
    \chi^\dagger 
    \,
    h_{\mathsf{BdG}} 
    \,
    \chi
    \,.
$ 
Here $\chi$ is the Nambu spinor spanning sublattice ($\sigma$), layer ($\tau$), 
and particle-hole spaces ($\mu$) and
\begin{equation}
    h_{\mathsf{BdG}} 
    = 
    h(\mathbf{r})
    \mu^3 
    + 
    \Delta (\mathbf{r}) \mu^1
    \,.
    \label{eq:hBdG}
\end{equation}
The Pauli matrices $\mu^i$ act on particle-hole space,
and $h(\mathbf{r})$ is the qBM Hamiltonian 
in Eq.~\eqref{eq:hqBM}. The pairing amplitude $\Delta(\mathbf{r})$ is inhomogeneous due to quasiperiodicity and obtained self-consistently by decoupling a local attractive interaction 
with coupling strength $U$, as detailed in Sec.~\ref{sec:SCDetails}. 

\begin{figure}[t!]
  \centering
  \includegraphics[width=0.48\textwidth]{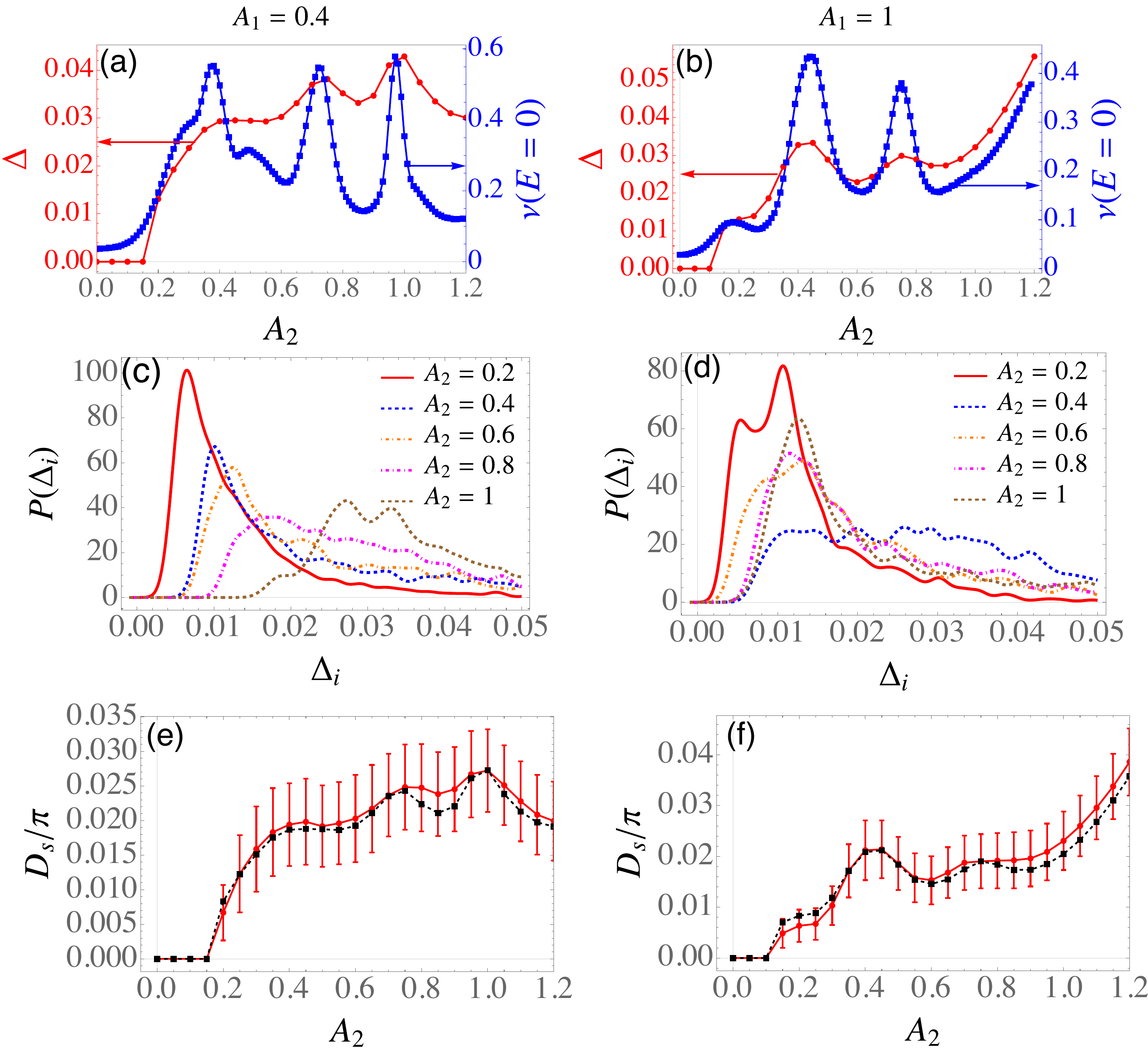}
  \caption{The average
  superconductivity order parameter $\Delta$, 
  normal-state zero-energy DOS $\nu(0)$, distribution of the local order parameter $\Delta_i$ and superfluid stiffness.
  (a): $\Delta$ and $\nu(E=0)$ along the cut $A_1=0.4$; (b): same as (a) along the cut $A_1=1$; (c): the distribution of local order parameter $\Delta_i$ along $A_1=0.4$; (d) same as (c) along the cut $A_1=1$; (e): superfluid stiffness (red) along $A_1=0.4$ (clean limit value $2\Delta/\pi$ indicated by black dashed curve); (f): same as (e) along $A_1=1$. Here $U=4v_F k_\theta$.}
  \label{fig:SC}
\end{figure}

Fig.~\ref{fig:SC} shows the average $\Delta$ 
and its distribution function
at half-filling obtained with KPM-based mean-field theory (Sec.~\ref{sec:KPMMFT}).
Figs.~\ref{fig:SC}(a) and (b) plot the average order parameter $\Delta$ along the two cuts $A_1=0.4$ and $1$, respectively.
Without the $A_2$ potential, the interaction strength is below the critical one 
for 2D Dirac fermions \cite{Uchoa2005, Uchoa2007, Rahul2013, IDP2014}
and $\Delta$ vanishes unless $A_1$ is 
close to the magic angles.
Unlike the BM model, superconductivity in the qBM model is not fine-tuned
and becomes 
nonzero as long as $A_2$ is not too weak.
The strength of the superconductivity 
correlates
with the change of 
the
DOS at 
the
Fermi energy,
peaking near filaments with maximal DOS.
However, the change in 
$\Delta$
is much shallower than that of the DOS. It results from the competition of the Fermi energy DOS and the spatial overlap of 
normal-state
wave functions. 
In BCS theory, 
superconductivity benefits from 
a larger
DOS at the Fermi energy.
Here however,
near the DOS peaks (the filaments), the spatial overlap of wave functions decays faster with energy difference than that in the lakes between filaments 
(not shown), 
which reduces the enhancement of superconductivity.
Figs.~\ref{fig:SC}(c) and (d) show the distribution of 
$\Delta_i$
and indicate that superconductivity is nonzero almost everywhere.

We compare $\Delta$ in the qBM and usual chiral BM models in Fig.~\ref{fig:SC_qBM_BM}. Fig.~\ref{fig:SC_qBM_BM}(b) shows $\Delta$ as a function of $U$ in the qBM model for a particular $A_{1,2}$ choice. The scaling indicates \emph{BCS-like} behavior $\Delta \sim \exp(-c/U)$, where $c$ is a constant, in the $U \rightarrow 0$ limit. This should be contrasted with the chiral BM model, where $\Delta = 0$ identically for sufficiently small $U$ away from the magic angle. In Sec.~\ref{sec:PoEE}, we show that \emph{finite-energy} SQHT states (present due to SWQC) and critical DOS scaling for class CI implies BCS scaling.  

Figs.~\ref{fig:SC}(e) and (f) show the superfluid stiffness 
along the two cuts (computed with a double-KPM method explained in Sec.~\ref{sec:KPMStiffness}), which closely follows the analytical prediction $2\Delta/\pi$ for a clean Dirac superconductor \cite{Kopnin08}. 
This value can also be anticipated from the optical conductivity of a dirty $s$-wave superconductor,
with imaginary part $\sigma_2(\omega) \simeq \pi \, \Delta \, \sigma_n / \omega$
\cite{Tinkham}. Since the normal-state
conductivity of the WZW Dirac model is $\sigma_n = 2 e^2/\pi^2$,
independent of disorder \cite{Evers08,Schnyder08}, we get  
$\sigma_2(\omega) \simeq e^2 \, 2 \Delta / \pi \omega$.
Superconductivity 
resists phase fluctuations and the transition temperature is determined by the strength of pairing amplitude.
Localizing Althsuler-Aronov effects have been found to suppress the stiffness in a quasiperiodic system \cite{Zhang22}; due to the effective topology these corrections are absent here \cite{Xie2015}.

\subsection{Discussion}

We have shown that the admixture of topology and quasiperiodicity in moir\'e materials can lead to remarkable quantum critical phenomena and robust superconductivity. Quasiperiodicity acts like disorder for strong mixing, but the coupling of same-valley Dirac fermions for small twists simulates topological surface states. 
Superconductivity is boosted
without magic-angle fine tuning \emph{and} without degrading the superfluid stiffness. 

Taking this toy model seriously, it is worth pointing out that in the microscopic TBLG lattice, our pairing is formally 
Fulde–Ferrell–Larkin–Ovchinnikov,
which would imply a pair-density wave on the lattice scale; further, it would be realized along with the pairing in the other valley to preserve time-reversal.
However, the chiral, quasiperiodic BM model we use in Eq.~\eqref{eq:hqBM} can be exactly constructed in a single layer of graphene with two mismatched $\sqrt{3}$-potentials, as was shown for one such potential in Ref.~\cite{Crepel2023}; in this system, pairing occurs between valleys (and is therefore, s-wave). 

The qBM model 
is a toy version of TTLG, where we expect distinct multifractality will be found \cite{Zhang25},
as well as 
in
twisted bilayers with a superimposed, incommensurate superlattice.
Topology plays an important role; while quasiperiodicity can drive transitions to critical states \cite{Devakul17}, it may not enhance and could even 
destroy
superconductivity \cite{Ticea24}. 
Nonetheless, this opens up the possibility for similar physics to occur in other 
moir\'e materials such as
transition-metal dichalcogenide 
multilayers.


\section{Numerical method for DOS, wave functions, and superconductivity \label{sec:KPM}}

The two kinds of 
Chebyshev 
orthogonal
polynomials are 
widely used to expand functions defined
in the interval $[-1, 1]$. Physical quantities of a 
condensed matter
system are usually
expressed as functions of the bounded Hamiltonian, and are 
in principle well-suited
to be expanded 
in terms of
Chebyshev polynomials. 
However,
the functions 
that determine
many interesting physical 
properties
such as 
the
spectral 
and
Green's functions 
are not smooth but have discontinuities or singularities. 
The truncated series expansion suffers from the so-called Gibbs phenomenon near 
discontinuities, where the expansion shows oscillation behavior.
The Gibbs oscillation of 
a
finite-series expansion cannot be resolved by simply increasing
the degree of
the
expansion. 
As a resolution, various kernels are
introduced to suppress the oscillation while keeping the series expansion
positive and normalized. This is the idea of the so-called kernel
polynomial method (KPM) \cite{KPM2006}. 
Here we give a brief review
of the KPM method, and related methods for wave functions, 
self-consistent mean-field theory, 
and 
superfluid stiffness.

\subsection{Basics on Chebyshev polynomials and the kernel polynomial method}

The Chebyshev polynomial of the first kind is 
most utilized
in KPM,
and its explicit expression is given by
\begin{equation}
T_{n}
\left( x \right)
=
\cos\left(n\arccos x\right)\,.
\end{equation}
The polynomials can be obtained recursively with
\begin{subequations}
\begin{align}
T_{0}\left(x\right) & =1\,,\\
T_{1}\left(x\right) & =x\,,\\
T_{n+1}\left(x\right) & =2xT_{n}\left(x\right)-T_{n-1}\left(x\right)\,.
\end{align}
\label{eq:recursiveTn}
\end{subequations}

A function $f\left(x\right)$
defined on the interval $x\in\left[-1,1\right]$ can be expanded as
\begin{align}
f\left(x\right) & =\frac{1}{\pi\sqrt{1-x^{2}}}\left[\mu_{0}+2\sum_{n=1}^{\infty}\mu_{n}T_{n}(x)\right]\,,\\
\mu_{n} & =\int_{-1}^{1}f\left(x\right)T_{n}\left(x\right)dx\,.
\end{align}
The Gibbs oscillation of the Chebyshev expansion can be suppressed
by introducing kernels
\begin{subequations}
\begin{align}
f\left(x\right) & =\frac{1}{\pi\sqrt{1-x^{2}}}\left[\tilde{\mu}_{0}+2\sum_{n=1}^{N_{\mathcal{C}}-1}\tilde{\mu}_{n}T_{n}\left(x\right)\right]\,,\\
\tilde{\mu}_{n} & =\mu_{n}g_{n}\,.
\end{align}
\end{subequations}
Here $N_{\mathcal{C}}$ is the 
maximum
degree of 
the
Chebyshev polynomials, 
$\mu_{n}$'s are the Chebyshev moments, and $g_{n}$'s are the kernel coefficients.
The Jackson kernel 
is commonly employed and suitable for most applications,
\begin{equation}
g_{n}
=
\frac{1}{N_{\mathcal{C}}+1}
\left[
\begin{aligned}
&\,
    \left(N_{\mathcal{C}}-n+1\right)\cos\left(\frac{\pi n}{N_{\mathcal{C}}+1}\right)
\\
    +
&\,
    \sin\left(\frac{\pi n}{N_{\mathcal{C}}+1}\right)
    \cot\left(\frac{\pi}{N_{\mathcal{C}}+1}\right)
\end{aligned}
\right]\,.
\label{eq:Jackson}
\end{equation}
In practice, we can obtain $f\left(x\right)$ at discrete points
using 
the
discrete Fourier transform. For example, we can choose 
\begin{equation}
    x_{k}
    =
    \cos\left(\frac{k+\frac{1}{2}}{N_{p}}\pi\right)
    \,,
    \qquad 
    k=0,1,\cdots,N_{p}-1\,.
\end{equation}
Here $N_{p}$ is the number of discrete points for $x$ and we can
choose $N_{p}\ge N_{\mathcal{C}}$ to utilize all the available Chebyshev
moments $\mu_{n}$. Then we have
\begin{equation}
    \gamma_{k}
    \equiv
    \pi\sqrt{1-x_{k}^{2}}
    \,
    f\left(x_{k}\right)
    =
    \tilde{\mu}_{0}
    +
    2\sum_{n=1}^{N_{\mathcal{C}}-1}
    \tilde{\mu}_{n}\cos\left[\frac{\pi n\left(k+\frac{1}{2}\right)}{N_{p}}\right]\!,
\end{equation}
which can be obtained simultaneously with
the
fast-Fourier-transform algorithm.
Another choice of $x$ discretization is given by
\begin{align}
    x_{k} 
    & =
    \cos\left(\frac{k\pi}{N_{p}-1}\right)\,,
    \qquad k=0,1,\cdots,N_{p}-1\,,\\
    \gamma_{k} 
    & =
    \pi\sqrt{1-x_{k}^{2}}
    \,
    f\left(x_{k}\right)
    =
    \tilde{\mu}_{0}
    +
    2\sum_{n=1}^{N_{\mathcal{C}}-1}\tilde{\mu}_{n}\cos\left(\frac{\pi nk}{N_{p}-1}\right)\,.
\end{align}
The advantage of the second one is that it 
accesses $x_{k}=0$ 
exactly
with odd $N_{p}$, which is desired when we are interested in the
scaling of 
the
zero-energy DOS. 
The two choices of $x_{k}$ corresponds
to the two kinds of Chebyshev-Gauss quadrature in numerical integration
of functions involving $f\left(x\right)$. 
With the first choice,
we have 
\begin{equation}
\int_{-1}^{1}f\left(x\right)g\left(x\right)dx
\approx
\frac{1}{N_{p}}
\sum_{k=0}^{N_{p}-1}
\gamma_{k}
\,
g\left(x_{k}\right)\,.
\end{equation}

\subsection{Density of states}

The density of states of a system can be obtained with KPM efficiently.
The global DOS of a system is given by
\begin{equation}
    \rho\left(E\right)=\frac{1}{V}\sum_{k}\delta\left(E-E_{k}\right)\,,
\end{equation}
with $V$ 
the volume of the system.
Here and below, we let
$\ket{i}$ and $\ket{k}$
denote the basis vectors in the
position and energy 
bases,
respectively.
In general, the spectrum of a system is bounded by $E_{\mathsf{min}}$
and $E_{\mathsf{max}}$, which can be obtained 
via the
Lanczos algorithm.
Then we can rescale the Hamiltonian,
\begin{equation}
\tilde{H}=aH+b\,,
\end{equation}
with 
\begin{align}
a & =\frac{2\left(1-\epsilon\right)}{E_{\mathsf{max}}-E_{\mathsf{min}}}\,,\\
b & =-\frac{\left(1-\epsilon\right)\left(E_{\mathsf{max}}+E_{\mathsf{min}}\right)}{E_{\mathsf{max}}-E_{\mathsf{min}}}\,,
\end{align}
 with $0<\epsilon\ll1$ to guarantee the numerical stability. Thus
we have 
\begin{equation}
\rho\left(E\right)=\frac{a}{V}\sum_{k}\delta\left(\tilde{E}-\tilde{E}_{k}\right)\,,
\end{equation}
 with $E=\left(\tilde{E}-b\right)/a$. The Chebyshev moments are then
given by
\begin{align}
    \mu_{n}
    =&\,
    \int_{-1}^{1}d\tilde{E}
    \,
    \rho\left(E\right)
    T_{n}\left(\tilde{E}\right)
\nonumber\\
    =&\,
    \frac{a}{V}
    \sum_{k}T_{n}\left(\tilde{E}_{k}\right)
    =
    \frac{a}{V}
    \mathsf{Tr}\left[T_{n}\left(\tilde{H}\right)\right]\,.
\end{align}
The trace can be efficiently evaluated by averaging over a few independent random vectors,
\begin{equation}
    \mu_{n}
    \approx
    \frac{a}{VR}
    \sum_{r=1}^R\bra{\psi_{r}}T_{n}\left(\tilde{H}\right)\ket{\psi_{r}}\,.
\end{equation}
Here $R$ is the number of random vectors 
$\left\{ \psi_{ri}\right\} $ satisfying
\begin{subequations}
\begin{align}
    \left\langle \psi_{ri}\right\rangle  & =0\,,\\
    \left\langle \psi_{ri} \, \psi_{r'j}\right\rangle  & =0\,,\\
    \left\langle \psi_{ri}^{*} \, \psi_{r'j}\right\rangle  & =\delta_{rr'}\delta_{ij}\,.
\end{align}
\end{subequations}
The angle brackets $\left\langle \cdots \right\rangle$ denote the average over the ensemble of random vectors. $\psi_{ri}$ is the $i$-th component of the $r$-th random vector, 
subject to some 
particular
distribution. In our numerics, we use 
the
uniform distribution for $\psi_{ri}$ or 
the
uniform distribution for $\theta$ with $\psi_{ri} = e^{i\theta}$.
The vector $\ket{\psi^{n}}\equiv T_{n}\left(\tilde{H}\right)\ket{\psi}$
can be obtained recursively using the 
relation in Eq.~\eqref{eq:recursiveTn},\begin{subequations}
\begin{align}
\ket{\psi^{0}} & =T_{0}\left(\tilde{H}\right)\ket{\psi}=\ket{\psi}\,,\\
\ket{\psi^{1}} & =T_{1}\left(\tilde{H}\right)\ket{\psi}=\tilde{H}\ket{\psi}\,,\\
\vdots &\phantom{=}\,\, \vdots \nonumber\\
\ket{\psi^{n+1}} & =T_{n+1}\left(\tilde{H}\right)\ket{\psi}=2\tilde{H}\ket{\psi^{n}}-\ket{\psi^{n-1}}\,.
\end{align}
\end{subequations}
Using the product relation of Chebyshev polynomials,
\begin{equation}
2T_{m}\left(x\right)T_{n}\left(x\right)=T_{m+n}\left(x\right)+T_{m-n}\left(x\right)\,.
\end{equation}
 Then we have 
\begin{align}
\mu_{2n} & =\frac{2a}{V}\braket{\psi^{n}|\psi^{n}}-\mu_{0}\,,\\
\mu_{2n+1} & =\frac{2a}{V}\braket{\psi^{n+1}|\psi^{n}}-\mu_{1}\,.
\end{align}
 The global DOS can be obtained with discrete Fourier transform of
$\tilde{\mu}_{n}=\mu_{n}g_{n}$. Typically, averaging a few ($\sim 10$) random vectors is enough to evaluate the global DOS accurately for large enough systems. As a result, the computational effort required for 
calculating the
global DOS is 
of the order of
$N N_\mathcal{C}$ for 
a
sparse Hamiltonian, with $N$ 
denoting
the dimension of the Hilbert space and $N_\mathcal{C}$ 
the degree of 
the
Chebyshev polynomials.

\subsection{Chebyshev filter methods for wave functions \label{sec:KPMWF}}

\subsubsection{Iterative filter method}
\label{subsec:Iterative-filter-method}

Various numerical methods based on 
the
KPM have been proposed to obtain the
eigenvalues and eigenvectors of a large sparse matrix. One of them is
the iterative Chebyshev filter diagonalization method proposed in
Ref.~\cite{Pieper2016}. The method aims to obtain the eigenvalue
and eigenvectors in a small interval $\left[E_{1},E_{2}\right]$.
It starts with an ensemble of random vectors, whose dimension is greater
than the number of states in the interval estimated from the DOS.
At each step of the iteration, the vectors are filtered with the kernel
polynomial expansion of a filter function. Then the set of filtered
vectors are orthogonalized to obtain an orthonormal basis in the interval.
The eigenvalues, eigenvectors and residual vectors are obtained with
Rayleigh-Ritz method. The steps can be repeated until the residual
of all the eigenpairs in the interval satisfy the preset criteria. Various
filter functions can be used for the method. One of them is the rectangular
function,
\begin{equation}
    W\left(x\right)=\Theta\left(x-E_{1}\right)\Theta\left(E_{2}-x\right)\,.
\end{equation}
A vector $\ket{\psi}$ can be filtered to amplify its components in
the interval via
\begin{equation}
    \ket{\psi^{\mathsf{F}}}
    =
    \sum_{n=0}^{N_{\mathcal{C}}}
    c_{n} 
    \,
    g_{n}
    \,
    T\left(\tilde{H}\right)
    \ket{\psi}\,.
\end{equation}
Here $g_{n}$ is the kernel and $c_{n}$ is 
an
expansion coefficient.
The latter
can be obtained exactly with
\begin{equation}
    c_{n}
    =
    \frac{2-\delta_{n0}}{\pi}
    \int_{-1}^{1}
    \frac{W\left(\tilde{x}\right)T_{n}\left(\tilde{x}\right)}{\sqrt{1-x^{2}}}
    dx\,.
\end{equation}

\subsubsection{Filter and shake method}

Instead of the iterative Chebyshev filter method, we use the filter
and shake method proposed in Ref.~\cite{Guan2021} to evaluate the
eigenvalues and eigenvectors in an interval. The method uses the so-called
exponential semicircle filter to obtain 
a vector $\ket{\psi_{0}}$
with numerically vanishing overlap with eigenvectors outside of the interval.
Then 
the 
vector 
is 
``shaken'' by Chebyshev iteration to obtain a set
of vectors, which are superpositions of eigenvectors in the desired
interval. The eigenvalues, eigenvectors and residual vectors are then
obtained via the Rayleigh-Ritz method using the orthonormal basis vectors. 

The exponential semicircle filter 
exploits
the fact that the Chebyshev
polynomials $T_n(x)$ diverge exponentially with $n$ when $x$ is outside
of the defining interval $\left[-1,1\right]$. When $\left|x\right|>1$,
we have
\begin{equation}
\arccos x=\begin{cases}
i\cosh^{-1}x & x>1\,,\\
\pi-i\cosh^{-1}x & x<-1\,.
\end{cases}
\end{equation}
Thus, we have 
\begin{align}
    T_{n}\left(x\right)
    =&\,
    \cos\left(n\arccos x\right)
    \nonumber\\   
    =&\,
    \begin{cases}
\cosh\left(n\cosh^{-1}x\right)\,, & x>1\,,\\
\left(-1\right)^{n}\cosh\left(n\cosh^{-1}x\right)\,, & x<-1\,.
\end{cases}
\end{align}
Using $\cosh^{-1}\left(1+x\right)\approx\sqrt{2x}$ with $x>0$, we
have
\[
\left|T_{n}\left(x\right)\right|\sim\frac{1}{2}e^{n\sqrt{2\left(\left|x\right|-1\right)}}\,,
\]
with $\left|x\right|>1$ and $n\sqrt{2\left(\left|x\right|-1\right)}\gg1$.
We can see that $\left|T_{n}\left(x\right)\right|$ grows exponentially
with the degree of Chebyshev iteration when $\left|x\right|>1$, while
$\left|T_{n}\left(x\right)\right|\ll1$ for $\left|x\right|<1$. Therefore,
the Chebyshev iteration itself provides an efficient filter by projecting
the desired energy window out of the defining range of the Chebyshev
polynomials. 

\begin{figure}
\centering

\includegraphics[width=0.8\columnwidth]{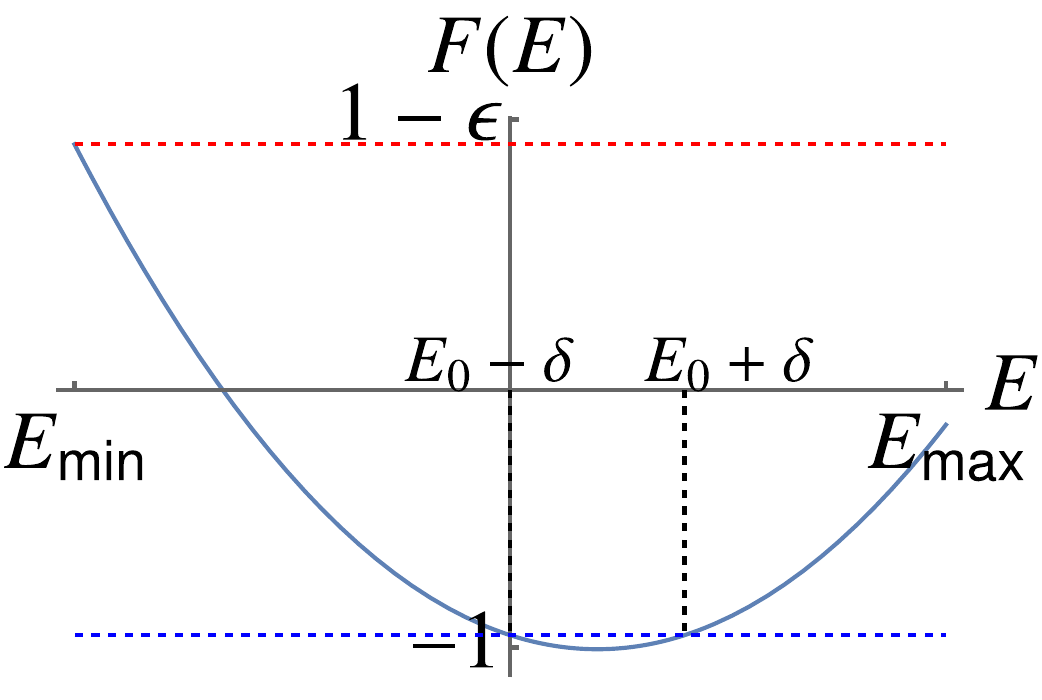}

\caption{Illustration of the exponential semicircle filter.}
\label{fig:ExpFilter}
\end{figure}

Consider a system with Hamiltonian $H$ and spectrum in 
the
range $\left[E_{\mathsf{min}},E_{\mathsf{max}}\right]$.
We seek the
eigenvalues and eigenstates in an energy window $\left[E_{0}-\delta,E_{0}+\delta\right]$.
We can use a parabolic function $F\left(H\right)$ {[}see Fig.~\ref{fig:ExpFilter}{]}
to project the energy spectrum of $H$ so that $F\left(E\right)>1$
when $E\in\left[E_{0}-\delta,E_{0}+\delta\right]$ and $F\left(E\right)<1$
when $E\notin\left[E_{0}-\delta,E_{0}+\delta\right]$. The parabolic
function $F\left(x\right)$ is given by
\begin{subequations}
\begin{align}
    F\left(H\right) 
    =&\,
    a^{2}\left(H-E_{0}\right)^{2}-c^{2}
\nonumber\\    
    =&\,\left[a\left(H-E_{0}\right)+c\right]\left[a\left(H-E_{0}\right)-c\right]\,,
\\
    a 
    =&\,
    \sqrt{\frac{2-\epsilon}{\mathcal{E}^{2}-\delta^{2}}}\,,\qquad c=\sqrt{\frac{\left(2-\epsilon\right)\delta^{2}}{\mathcal{E}^{2}-\delta^{2}}+1}\,,
\\
    \mathcal{E} 
    =&\,
    \max\left\{ \left|E_{\mathsf{min}}-E_0\right|,\left|E_{\mathsf{max}}-E_{0}\right|\right\} \,.
\end{align}
\end{subequations}

Performing Chebyshev iteration on $F\left(H\right)$ on a random 
state
$\ket{\psi}$, we have 
\begin{subequations}
\begin{align}
\ket{\psi^{0}} & =\ket{\psi}\,,\\
\ket{\psi^{1}} & =F\left(H\right)\ket{\psi^0}\,,\\
\cdots\nonumber\\
\ket{\psi^{n}} & =2F\left(H\right)\ket{\psi^{n}}-\ket{\psi^{n-1}}\,.
\end{align}
\end{subequations}
 The initial random vector can be written as
\begin{equation}
\ket{\psi}=\sum_{m=1}^N c_{m}\ket{m}\,,
\end{equation}
with $\ket{m}$ 
denoting
the eigenstate of $H$ with energy $E_{m}$
and $c_{m}$ 
a
random number with $\left|c_{m}\right|\sim\frac{1}{\sqrt{N}}$.
At order $n$ of the Chebyshev iteration, we have
\begin{subequations}
\begin{align}
\ket{\psi^{n}} & =\sum_{m}\tilde{c}_{m}\ket{m}\,,\\
\tilde{c}_{m} & =c_{m}T_{n}\left[F\left(E_{m}\right)\right]\,.
\end{align}
\end{subequations}
When $E_{m}$ falls out of the interval $\left[E_{0}-\delta,E_{0}+\delta\right]$,
$T_{n}=\cos\left[n\arccos F\left(E_{m}\right)\right]\in\left[-1,1\right]$
and $\tilde{c}_{m}$ does not increase with the degree of Chebyshev
expansion $n$. However, when $E_{m}\in\left[E_{0}-\delta,E_{0}+\delta\right]$,
we have 
\begin{align}
    T_{n}\left[F\left(E_{m}\right)\right] 
    & =
    \left(-1\right)^{n}
    \cosh\left[n\cosh^{-1}F\left(E_{m}\right)\right]
    \nonumber \\
    & \approx
    \left(-1\right)^{n}
    \frac{1}{2} \,
    e^{n\sqrt{2\left[-a^{2}\left(E_{m}-E_{0}\right)^{2}+c^{2}-1\right]}}
    \nonumber \\
    & =
    \frac{\left(-1\right)^{n}}{2}
    e^{\frac{2n}{\sqrt{\mathcal{E}^{2}-\delta^{2}}}\sqrt{\delta^{2}-\left(E_{m}-E_{0}\right)^{2}}}\,.
\end{align}
 Thus $\tilde{c}_{m}$ grows as the exponential of a semicircle when
$E_{m}\in\left[E_{0}-\delta,E_{0}+\delta\right]$ and $n$ is large.
Focusing on the interval within the half-width of the semi-circle,
we have 
\begin{equation}
    \left|\frac{\tilde{c}_{m}}{c_{m}}\right|>\frac{1}{2}e^{\frac{n\delta}{\mathcal{E}}}\,.
\end{equation}
 It turns out that the 
exponential semicircle
filter is much more efficient than the rectangular
one
and it can be used to accelerate the iterative method in 
Sec.~\ref{subsec:Iterative-filter-method}. 
Moreover, Ref.~\cite{Guan2021}
proposed that a complete basis for the states in the desired interval
can be obtained by shaking the filtered state with Chebyshev iteration.
In numerics, a double-precision float number has about $16$ significant
figures. When $n>\frac{\mathcal{E}}{\delta}\log\left(10^{16}\right)$,
the $\ket{\psi^{n}}$ contains numerically vanishing 
components
of
eigenstates
outside of the desired interval and 
\begin{equation}
    \ket{\psi^{n}}
    \approx
    \sum_{E_{0}-\delta<E_{m}<E_{0}+\delta}\tilde{c}_{m}\ket{m}\,.
\end{equation}
Now we take Chebyshev iteration with scaled Hamiltonian $\tilde{H}$
and $\ket{\psi^{n}}$ as initial state. We have 
\begin{subequations}
\begin{align}
\ket{\phi^{0}} & =\ket{\psi^{n}}\,,\\
\ket{\phi^{1}} & =\tilde{H}\ket{\phi^{0}}\,,\\
\cdots\nonumber\\
\ket{\phi^{l+1}} & =2\tilde{H}\ket{\phi^{l}}-\ket{\phi^{l-1}}\,.
\end{align}
\end{subequations}
Then
\begin{equation}
    \ket{\phi^{l}}\approx\sum_{E_{0}-\delta<E_{m}<E_{0}+\delta}\tilde{c}_{m}\cos\left(l\arccos\tilde{E}_{m}\right)\ket{m}\,.
\end{equation}
Thus the amplitude of $\ket{m}$ in the $\ket{\phi^{l}}$ oscillates
with its energy and the phase is given by 
\begin{equation}
    l\arccos\tilde{E}_{m}
    \approx 
    l\arccos{\tilde{E}_0}-\frac{l\left(\tilde{E_m}-\tilde{E_0}\right)}{\sqrt{1-\tilde{E}_0^2}}\,,
\end{equation}
with $\left|\tilde{E_m}-\tilde{E_0}\right|\ll1$.
The phases of the states in the desired interval now distribute across the interval
$ l\arccos{\tilde{E}_0}-\frac{l}{\sqrt{1-\tilde{E}_0^2}}\left[-\tilde{\delta}, \tilde{\delta}\right]$ with $\tilde{\delta}\ll 1$.
When 
$l \tilde{\delta}/{\sqrt{1-\tilde{E}_0^2}} \approx\frac{\pi}{2}\mod\pi$, the phases of the all the states in the desired interval runs across an interval of length $\pi$. If the energies of the states uniformly distributed in the desired interval, the phases are also uniformed distributed in the length $\pi$ interval around $l\pi/2$ if $\tilde{E_0}\approx 0$. Thus for each $l$, we have a different superposition of eigenstates in the desired interval. 
Thus we can keep one state $\ket{\phi^{l}}$ for the basis for every
$\lfloor\frac{\pi\sqrt{1-\tilde{E}_0^2}}{2\tilde{\delta}}\rfloor$ Chebyshev iteration.
Moreover, observing that $\arccos\tilde{E}_{m}\approx\frac{\pi}{2}+\tilde{E}$
when $\tilde{E}_{m}\approx0$ and we have 
\begin{align}
    \cos\left(l\arccos\tilde{E}_{m}\right)
    \approx&\,
    \cos\left(\frac{\pi l}{2}+l\tilde{E}\right)
\nonumber\\
    =&\,
    \begin{cases}
\left(-1\right)^{\frac{l}{2}}\cos\left(l\tilde{E}\right) & l\in\text{even}\,,\\
\left(-1\right)^{\frac{l+1}{2}}\sin\left(l\tilde{E}\right) & l\in\text{odd}\,.
\end{cases}
\end{align}
Thus we can keep both $\ket{\phi^{l}}$ and $\ket{\phi^{l+1}}$ in
the basis when $l=\lfloor\frac{j\pi\mathcal{E}}{2\delta}\rfloor$.
Shaking $\ket{\psi^{n}}$ with Chebyshev iteration, we can generate
a basis for the states in the desired interval. The eigenvalue and
eigenstates can then be generated with this basis via the Rayleigh-Ritz
method. 

Now we summarize the filter and shake method as follows:
\begin{enumerate}
\item Calculate $E_{\mathsf{min}}$ and $E_{\mathsf{max}}$ of $H$ with
the Lanczos method.
\item Calculate the DOS of the system using KPM and estimate the number
of states $N_{T}$ in the desired interval $\left[E_{0}-\delta,E_{0}+\delta\right]$.
\item Generate the filtered state $\ket{\psi^{n}}$ from a random state
using the exponential semicircle filter and $n>\frac{\mathcal{E}}{\delta}\log\left(10^{16}\right)$
with $\mathcal{E}=\max\left\{ \left|E_{\mathsf{min}}-E_{0}\right|,\left|E_{\mathsf{max}}-E_{0}\right|\right\} $.
\item Shake $\ket{\psi^{n}}$ with Chebyshev iteration and $\ket{\phi^{0}}=\ket{\psi^{n}}$.
Keep $\ket{\phi^{l}}$ and $\ket{\phi^{l+1}}$ to the basis with $\frac{l\delta}{\mathcal{E}}\approx\frac{\pi}{2}$ (assuming $E_0\approx 0$)
and construct a basis $\left\{ \Phi_{j}\right\} $ containing $N_{S}>N_{T}$
states. $N_{S}=1.5N_{T}$ or $N_{S}=2N_{T}$ is usually sufficient.
\item Orthogonalize the set of basis vector $\left\{ \Phi_{j}\right\} $
and obtain an orthonormal basis containing $N_{B}\le N_{S}$ states.
This can be done with the SVQB algorithm.
\item Calculate the eigenvalue, eigenvectors and residual vectors using
the orthonormal basis via Rayleigh-Ritz method. 
\item Keep the states with $E\in\left[E_{0}-\delta,E_{0}+\delta\right]$
and residual smaller than the preset criteria $\eta$.
\end{enumerate}
The method works best when the eigenvalues distribute uniformly
across the desired interval. When there are degeneracies, we should
generate a number of filtered states $\ket{\psi^{n}}$ with independent
random initial vectors and generate a subset of the basis for each
filtered state. When the number of filtered states is greater than the degeneracy of states in the interval, the filter and shake method works well. Moreover, the method is more efficient by using an appropriate number of filtered states, 
balancing the computational time for the filter step and the shake step.

\subsection{Self-consistent BCS theory based on KPM \label{sec:KPMMFT}}

For 
an
inhomogeneous system with interactions, mean-field order parameters
also become inhomogeneous. 
The 
spatially varying
order parameter 
generally
prohibits
analytical evaluation and exact diagonalization (ED) is often required
to perform the calculation. However, it takes a lot of memory and
computational time to diagonalize a large Hamiltonian matrix and the
system size accessible is limited at the order of $10^{4}$ lattice
sites. The KPM based self-consistent mean field theory allows us to
access larger systems with much less cost of memory and computational
time.

\subsubsection{Self-consistent BdG theory for a lattice model}

Now we consider a lattice model with attractive Hubbard interaction,
\begin{equation}
    H=-t\sum_{\left\langle ij\right\rangle \sigma}\left(c_{i\sigma}^{\dagger}c_{j\sigma}+H.c.\right)-U\sum_{i}n_{i\uparrow}n_{i\downarrow}-\mu\sum_{i}n_{i}\,.
\end{equation}
It can be solved by the mean-field ansatz,
\begin{equation}
\Delta_i=-U\left\langle c_{i\downarrow}c_{i\uparrow}\right\rangle \,,\qquad\left\langle n_{i\sigma}\right\rangle =\left\langle c_{i\sigma}^{\dagger}c_{i\sigma}\right\rangle =\frac{1}{2}\left\langle n\right\rangle \,.
\end{equation}
The BdG mean-field Hamiltonian can be written as
\begin{align}
    H_{\mathsf{eff}}
    =&\,
    -t\sum_{\left\langle i,j\right\rangle \sigma}\left(c_{i\sigma}^{\dagger}c_{j\sigma}+H.c.\right)
    +\sum_{i}\left(V_{i}-\tilde{\mu}_{i}\right)n_{i\sigma}
\nonumber\\
&\,
    +\sum_{i}\left(\Delta_{i}c_{i\uparrow}^{\dagger}c_{i\downarrow}^{\dagger}+\Delta_{i}^{*}c_{i\uparrow}c_{i\downarrow}\right)\,.
\end{align}
Here $\tilde{\mu} = \mu + U\left<n\right>/2$ 
incorporates the Hartree shift.
$H_{\mathsf{eff}}$ can be diagonalized by the Bogoliubov transformation,
\begin{align}
    c_{i\uparrow} 
    =&\,
    \sum_{n}'\left[u_{n}^{i}\alpha_{n}-\left(v_{n}^{i}\right)^{*}\beta_{n}^{\dagger}\right]\,,
\\
    c_{i\downarrow}
    =&\,
    \sum_{n}'\left[\left(v_{n}^{i}\right)^{*}\alpha_{n}^{\dagger}+u_{n}^{i}\beta_{n}\right]\,,
\\
\left[\begin{array}{c}
c_{\uparrow}
\\
\left(c_{\downarrow}^{\dagger}\right)^{\mathsf{T}}
\end{array}\right] 
=&\,
\left[\begin{array}{cc}
u & -v^{*}\\
v & u^{*}
\end{array}\right]
\left[\begin{array}{c}
\alpha\\
\left(\beta^{\dagger}\right)^{\mathsf{T}}
\end{array}\right]\,,
\end{align}
with $\alpha$ and $\beta{}^{\dagger}$ quasiparticle operators, $\alpha_{i}$
and $\beta_{i}$ correspond to quasiparticles with the same energy
and $\sum'$ sums over all the positive energy states. $u,v$ 
denote $N \times N$ matrices
and $\left(u_{1n},\cdots,u_{Nn},v_{1n},\cdots,v_{Nn}\right)^{\mathsf{T}}$
is
the $n$-the positive energy $E_{n}$ eigenstate of the $2N\times2N$
BdG Hamiltonian,
\begin{equation}
\left[\begin{array}{cc}
\hat{h}_{0} & \Delta\\
\Delta^{\dagger} & -\hat{h}_{0}^{\mathsf{T}}
\end{array}\right]\left[\begin{array}{c}
u_{n}\\
v_{n}
\end{array}\right]=E_{n}\left[\begin{array}{c}
u_{n}\\
v_{n}
\end{array}\right]\,,
\end{equation}
 with $u_{n}=\left(u_{1n},\cdots,u_{Nn}\right)^{\mathsf{T}}$ and
$v_{n}=\left(v_{1n},v_{2n},\cdots,v_{Nn}\right)^{\mathsf{T}}$, 
$(\hat{h}_0)_{ij}=-t\left(\delta_{i,j+1}+\delta_{i+1,j}\right)-t\left(\delta_{i1}\delta_{jN}+\delta_{iN}\delta_{j1}\right)+\left(V_{i}-\tilde{\mu}_{i}\right)\delta_{ij}$
and $\Delta_{ij}=\Delta_{i}\delta_{ij}$. Thus we have 
\begin{equation}
H_{\mathsf{eff}}=\sum_{n}'E_{n}\left(\alpha_{n}^{\dagger}\alpha_{n}-\beta_{n}\beta_{n}^{\dagger}\right)\,.
\end{equation}
The self-consistent conditions at $T=0$ are then given by
\begin{equation}
\Delta_{i}=U\sum_{n}'u_{n}^{i}\left(v_{n}^{i}\right)^{*}\,,\qquad\left\langle n_{i}\right\rangle =2\sum_{n}'\left|v_{n}^{i}\right|^{2}\,.
\end{equation}
At finite temperature, we have
\begin{align}
    \Delta_{i} 
    =&\,
    -U\left\langle c_{i\downarrow}c_{i\uparrow}\right\rangle 
\nonumber\\
    =&\,
    -U\sum_{mn}'\left\langle \left[\left(v_{m}^{i}\right)^{*}\alpha_{m}^{\dagger}+u_{n}^{i}\beta_{m}\right]\left[u_{n}^{i}\alpha_{n}-\left(v_{n}^{i}\right)^{*}\beta_{n}^{\dagger}\right]\right\rangle 
\nonumber\\
    =&\,
    U\sum_{n}'u_{n}^{i}\left(v_{n}^{i}\right)^{*}\left[1-2f\left(E_{n}\right)\right]\,
\end{align}
and
\begin{align}
\left\langle n_{i}\right\rangle  & =\left\langle c_{i\uparrow}^{\dagger}c_{i\uparrow}\right\rangle +\left\langle c_{i\downarrow}^{\dagger}c_{i\downarrow}\right\rangle \nonumber \\
 & =2\sum_{n}'\left[\left|u_{n}\right|^{2}f\left(E_{n}\right)+\left|v_{n}^{i}\right|^{2}\left(1-f\left(E_{n}\right)\right)\right]\,.
\end{align}
 Here $f\left(E_{n}\right)=\frac{1}{e^{\beta E_{n}}+1}$ is the Fermi
distribution function for the quasiparticles with energy $E_{n}$. 

It turns out to be convenient to write the eigenstates in the following
form,
\begin{align}
\left(\begin{array}{cc}
u & -v^{*}\\
v & u^{*}
\end{array}\right)
=&\,
\left(\begin{array}{cccccc}
u_{1}^{i} & \cdots & u_{N}^{i} & -\left(v_{1}^{i}\right)^{*} & \cdots & -\left(v_{N}^{i}\right)^{*}\\
v_{n}^{i} & \cdots & v_{N}^{i} & \left(u_{1}^{i}\right)^{*} & \cdots & \left(u_{N}^{i}\right)^{*}
\end{array}\right)
\nonumber\\
=&\,
\left(\begin{array}{cccccc}
u_{1}^{i} & \cdots & u_{N}^{i} & u_{-1}^{i} & \cdots & u_{-N}^{i}\\
v_{n}^{i} & \cdots & v_{N}^{i} & v_{-1}^{i} & \cdots & v_{-1}^{i}
\end{array}\right)\,,
\end{align}
with the first $N$ columns for the $N$ positive energy states and
the last $N$ columns for the negative energy states. Denote the eigenstate
matrix as
\begin{equation}
\left(\begin{array}{cccccc}
u_{1}^{i} & \cdots & u_{N}^{i} & u_{-1}^{i} & \cdots & u_{-N}^{i}\\
v_{n}^{i} & \cdots & v_{N}^{i} & v_{-1}^{i} & \cdots & v_{-1}^{i}
\end{array}\right)\,,
\end{equation}
 with $u_{-n}^{i}=-\left(v_{n}^{i}\right)^{*}$ and $v_{-n}^{i}=\left(u_{n}^{i}\right)^{*}$.
Here 
\begin{gather}
u_{n}^{i}=\braket{i\uparrow|n}\,,
\qquad 
v_{n}=\braket{i\downarrow|n}\,,
\qquad
1\le i\le N,
\nonumber\\
-N\le n\le N\,,
\qquad
n\neq0\,.
\end{gather}
Here $\ket{i\uparrow}=c_{i\uparrow}^{\dagger}\ket{0}$ and $\ket{i\downarrow}=c_{i\downarrow}\ket{0}$
with $\ket{0}$ being the BCS ground state without quasiparticle excitations.
The corresponding 
Bogoliubov
transformation is given by 
\begin{align}
    c_{i\uparrow} 
    =&\,
    \sum_{n}'\left(u_{n}^{i}\alpha_{n}+u_{-n}^{i}\alpha_{-n}^{\dagger}\right)\,,
\\
    c_{i\downarrow}
    =&\,
    \sum_{n}'\left[\left(v_{n}^{i}\right)^{*}\alpha_{n}^{\dagger}+\left(v_{-n}^{i}\right)^{*}\alpha_{-n}\right]\,,
\\
\left[\begin{array}{c}
c_{\uparrow}\\
\left(c_{\downarrow}^{\dagger}\right)^{\mathsf{T}}
\end{array}\right] 
=&\,
\left[\begin{array}{cc}
u_{+} & u_{-}\\
v_{+} & v_{-}
\end{array}\right]
\left[\begin{array}{c}
\alpha\\
\left(\alpha^{\dagger}\right)^{\mathsf{T}}
\end{array}\right]\,.
\end{align}
In summary, we have
\begin{subequations}
\begin{align}
\Delta_{i} & =-U\sum_{n}u_{n}^{i}\left(v_{n}^{i}\right)^{*}f\left(E_{n}\right)\,,\qquad E_{-n}=-E_{n}\,,\\
n_{i} & =2\sum_{n}\left|u_{n}^{i}\right|^{2}f\left(E_{n}\right)\,,\\
u_{n}^{i} & =\braket{i\uparrow|n}\,,\qquad v_{n}=\braket{i\downarrow|n}\,.
\end{align}
\end{subequations}
These expressions for $\Delta_{i}$ and $n_{i}$
are easier to implement at finite temperature and are suitable to
be generalized to the KPM method.

\subsubsection{KPM Implementation of the self-consistent BdG theory}

The BdG Hamiltonian has particle-hole symmetry and its spectrum is
symmetric with respective to $0$. Thus we can rescale the BdG Hamiltonian
with 
\begin{equation}
\tilde{H}=aH\,,\qquad a=\frac{1-\epsilon}{\left|E_{g}\right|}\,,
\end{equation}
with $E_{g}$ being the smallest eigenenergy of the BdG Hamiltonian. 

We define the pairing and density spectral functions,
\begin{align}
    \Delta_{i}\left(E\right) 
    =&\,
    -U\sum_{n}u_{n}^{i}\left(v_{n}^{i}\right)^{*}\delta\left(E-E_{n}\right)
\nonumber\\    
    =&\,
    -U\sum_{n}\braket{i\uparrow|n}\braket{n|i\downarrow}\delta\left(E-E_{n}\right)\,,
\\
    \rho_{i}\left(E\right) 
    =&\,
    2\sum_{n}\left|u_{n}^{i}\right|^{2}\delta\left(E-E_{n}\right)
\nonumber\\    
    =&\,
    2\sum_{n}\braket{i\uparrow|n}\braket{n|i\uparrow}\delta\left(E-E_{n}\right)\,.
\end{align}
 Therefore, the local pairing amplitude and local density can be obtained
via
\begin{align}
    \Delta_{i} 
    & =
    \int dE
    \,
    \Delta_{i}\left(E\right)f\left(E\right)\,,
    \\
    n_{i} 
    & =
    \int dE
    \,
    \rho_{i}\left(E\right)f\left(E\right)\,.
\end{align}
The rescaled spectral functions of the scaled Hamiltonian are given
by
\begin{align}
    \tilde{\Delta}_{i}\left(\tilde{E}\right) 
    =&\,
    -U\sum_{n}\braket{i\uparrow|n}\braket{n|i\downarrow}\delta\left(\tilde{E}-\tilde{E}_{n}\right)
\nonumber\\    
    =&\,
    \frac{1}{a}\Delta_{i}\left(E=a\tilde{E}\right)\,,
\\
    \tilde{\rho}_{i}\left(\tilde{E}\right) 
    =&\,
    2\sum_{n}\braket{i\uparrow|n}\braket{n|i\uparrow}\delta\left(\tilde{E}-\tilde{E}_{n}\right)
\nonumber\\    
    =&\,
    \frac{1}{a}\rho_{i}\left(E=a\tilde{E}\right)\,.
\end{align}
The Chebyshev moments of the rescaled spectral functions are given
by 
\begin{align}
    \mu_{m}^{\Delta_{i}} 
    =&\,
    \int_{-1}^{1}\tilde{\Delta}_{i}\left(\tilde{E}\right)T_{m}\left(\tilde{E}\right)d\tilde{E}
\nonumber\\    
    =&\,
    -U\sum_{n}\braket{i\uparrow|n}\braket{n|i\downarrow}T_{m}\left(\tilde{E}_{n}\right)
\nonumber \\
    =&\,
    -U\left(\bra{i\downarrow}T_{m}\left(\tilde{H}\right)\ket{i\uparrow}\right)^{*}\,,
\\
    \mu_{m}^{\rho_{i}} 
    =&\, 
    \int_{-1}^{1}\tilde{\rho}_{i}\left(\tilde{E}\right)T_{m}\left(\tilde{E}\right)d\tilde{E}
\nonumber\\
    =&\,
    2\sum_{n}\braket{i\uparrow|n}\braket{n|i\uparrow}T_{m}\left(\tilde{E}_{n}\right)
\nonumber \\
    =&\,
    2\bra{i\uparrow}T_{m}\left(\tilde{H}\right)\ket{i\uparrow}\,.
\end{align}
The Chebyshev iteration can be performed starting with $\ket{i\uparrow}$,
\begin{subequations}
\begin{align}
\left|\psi_{0}\right> & =\left|i\uparrow\right>\,,\\
\left|\psi_{1}\right> & =\tilde{H}\left|\psi_0\right>\,,\\
\cdots\nonumber\\
\left|\psi_{m+1}\right> & =2\tilde{H}\left|\psi_{m}\right>-\left|\psi_{m-1}\right>\,.
\end{align}
\end{subequations}
Thus we have the Chebyshev moments,
\begin{align}
\mu_{m}^{\Delta_{i}} & =-U\braket{i\downarrow|\psi_{m}}\,,\\
\mu_{m}^{\rho_{i}} & =2\braket{i\uparrow|\psi_{m}}\,.
\end{align}
The Chebyshev moments can be obtained by reading spin up and down
components of $\ket{\psi_{m}}$ without performing the inner product. 

With $N_{\mathcal{C}}$ Chebyshev moments of $\Delta_{i}$ and $\mu_{i}$
available, we can evaluate the scaled spectral functions at points,
\begin{equation}
    \tilde{E}_{k}
    =
    \cos\left[\frac{\pi\left(k+\frac{1}{2}\right)}{N_{\mathsf{p}}}\right]\,,\qquad k=0,1,\cdots,N_{\mathsf{p}}-1\,.
\end{equation}
 Usually we consider $N_{p}\ge N_{\mathcal{C}}$ (e.g. $N_{p}=2N_{\mathcal{C}}$)
and $\tilde{\Delta}_{i}\left(\tilde{E}_{k}\right)$ can be evaluated
through the discrete cosine transform from the 1D array 
\[
    \left\{ \tilde{\mu}_{0},\tilde{\mu}_{1},\cdots,\tilde{\mu}_{N_{\mu}-1}\underset{N_{p}-N_{\mathcal{C}}}{,\underbrace{0,\cdots,0}}\right\},
\]
Then we have,
\begin{align}
    \gamma_{k}^{\Delta_{i}} 
    =&\,
    \pi\sqrt{1-\tilde{E}_{k}^{2}}\tilde{\Delta}_{i}\left(\tilde{E}_{k}\right)
\nonumber\\
    =&\,
    \tilde{\mu}_{0}^{\Delta_{i}}+2\sum_{m=1}^{N_{\mathcal{C}}-1}\tilde{\mu}_{n}^{\Delta_{i}}
    \cos\left[\frac{\pi n\left(k+1/2\right)}{N_{p}}\right]\,,
\\
    \gamma_{k}^{\rho_{i}} 
    =&\,
    \pi\sqrt{1-\tilde{E}_{k}^{2}}\tilde{\rho}_{i}\left(\tilde{E}_{k}\right)
\nonumber\\
    =&\,
    \tilde{\mu}_{0}^{\rho_{i}}+2\sum_{m=1}^{N_{\mathcal{C}}-1}\tilde{\mu}_{n}^{\Delta_{i}}
    \cos\left[\frac{\pi n\left(k+1/2\right)}{N_{p}}\right]\,.
\end{align}
 Here 
\begin{equation}
\tilde{\mu}_{m}=g_{m}\mu_{m}\,,
\end{equation}
 with $g_{m}$ being the Jackson kernel in Eq.~\eqref{eq:Jackson}.
The integrals involving $f\left(x\right)$ can be obtained via Chebyshev-Gauss
quadrature,
\begin{align}
    \int_{-1}^{1}f\left(x\right)g\left(x\right)dx 
    =&\,
    \int_{-1}^{1}dx\frac{\sqrt{1-x^{2}}f\left(x\right)g\left(x\right)}{\sqrt{1-x^{2}}}
\nonumber\\
    =&\,
    \frac{1}{N_{p}}\sum_{k=0}^{N_{p}-1}\gamma_{k}g\left(x_{k}\right)\,.
\end{align}
 Thus we have 
\begin{align}
    \Delta_{i} 
    =&\,
    \int_{-E_{\mathsf{max}}}^{E_{\mathsf{max}}}\Delta_{i}\left(E\right)f\left(E\right)dE
\nonumber\\    
    =&\,\int_{-1}^{1}\tilde{\Delta}_{i}\left(\tilde{E}\right)f\left(\tilde{E}\right)d\tilde{E}=\frac{1}{N_{p}}\sum_{k=0}^{N_{p}-1}\gamma_{k}^{\Delta_{i}}f\left(E_{k}\right)\,,
\\
    n_{i} 
    =&\,
    \int_{-E_{\mathsf{max}}}^{E_{\mathsf{max}}}\rho_{i}\left(E\right)f\left(E\right)dE=\int_{-1}^{1}\tilde{\rho}_{i}\left(\tilde{E}\right)f\left(\tilde{E}\right)d\tilde{E}
\nonumber\\    
    =&\,
    \frac{1}{N_{p}}\sum_{k=0}^{N_{p}-1}\gamma_{k}^{\rho_{i}}f\left(E_{k}\right)\,.
\end{align}
 Here $f\left(E_{k}\right)$ is the Fermi distribution function. 

The self-consistent BdG theory can be performed until the difference
of the output order parameters $\left\{ \Delta_{i},n_{i}\right\} $
are close enough to the input ones. At each iteration, $\Delta_{i}$
and $n_{i}$ need to be evaluated site by site. The convergence of
the iteration can be further accelerated by 
Broyden's method \cite{Sriv1984,Johnson1988}.
The KPM-based method requires only a small amount of memory and computation
effort that scales with $N^{2}N_{\mathcal{C}}$, with $N$ being
the number of lattice sites. On the other hand, the computational
effort of ED scales with $N^{3}$. On our test platform with $2$
CPUs [Intel(R) Xeon(R) CPU E5-2650 v4 @ 2.20GHz], the KPM-based method
surpasses the ED-based one when $N\approx2000$ and $N_{\mathcal{C}}=4096$.
Therefore, the KPM-based mean-field method is superior 
to
ED-based one as long as we need to deal with systems with more than several
thousand lattice sites. Moreover, the KPM-based method can be easily
accelerated with a parallel computation scheme like MPI. The communication
cost is very low and the acceleration scales linearly with the number
of MPI processes. Thus, very large system sizes can be accessed with
enough resources.

\subsubsection{Continuum model with local interaction}

For the qBM-BCS model, the Hamiltonian can be written as
\begin{equation}
    H
    =
    \int d\mathbf{r}
    \,
    \chi^{\dagger}
    \left[
        h\left(\mathbf{r}\right)
        \mu^{3}
        +
        \Delta\left(\mathbf{r}\right)
        \mu^{1}
    \right]\chi\,,
\end{equation}
where $h\left(\mathbf{r}\right)$ is the qBM one-body Hamiltonian
from Eq.~(\ref{eq:hqBM}).
Both terms are local in real space, but only the second term can
be expressed as a sparse matrix in 
position space.
The first term
involves the Dirac operator $-i\boldsymbol{\sigma}\cdot\nabla$, which
cannot be expressed as a short-range lattice Hamiltonian without introducing
the fermion-doubling problem. Therefore, we work 
simultaneously with both real- and momentum-space lattices,
switching back and forth as necessary.
Starting with a real-space
state $\ket{\psi}$, we have 
\begin{equation}
H\ket{\psi}=H_{2}\ket{\psi}+\mathcal{F}^{-1}\left(H_{1}\ket{\mathcal{F}\psi}\right)\,.
\end{equation}
Here $\mathcal{F}$ is the Fourier transform from real space to momentum
space, and 
\begin{align}
    H_{1} 
    & =
    \int d\mathbf{r}
    \,
    \chi^{\dagger}
    h\left(\mathbf{r}\right)\mu^{3}\chi
    =
    \intop_{\mathbf{k}_{1}\mathbf{k}_{2}}
    \chi_{\mathbf{k}_{1}}^{\dagger}
    \hat{h}_{\mathbf{k}_{1}\mathbf{k}_{2}}
    \chi_{\mathbf{k}_{2}}\,,
    \\
    H_{2} 
    & =
    \intop d\mathbf{r}
    \,
    \chi^{\dagger}
    \Delta\left(\mathbf{r}\right)
    \mu^{1}
    \chi\,.
\end{align}
The Hamiltonian matrix of the first term is sparse in momentum space
and the action of $H_{1}$ on the momentum space wave function $\ket{\mathcal{F}\psi}$
is cheap
to calculate.
On the other hand, the Hamiltonian matrix 
of the
second term is
sparse in real space, 
as it acts locally
on the real-space wave function.
With this procedure, we can deal with a system with Dirac cones exactly
without involving the fermion-doubling problem. The most computational
challenging part now lies on the Fourier transform of wave functions,
which can be done by fast-Fourier algorithm provided in vendor software packages.\\

\subsection{Calculation of superfluid stiffness with KPM \label{sec:KPMStiffness}}

\subsubsection{Superfluid stiffness of lattice model}

Consider the monochromatic electromagnetic field with vector potential
$\mathbf{A}\left(\mathbf{r}_{i},t\right)$,
\begin{equation}
\mathbf{A}\left(\mathbf{r}_{i},t\right)=\Re\left[\mathbf{A}\left(\omega,\mathbf{q}\right)e^{-i\omega t+i\mathbf{q}\cdot\mathbf{r}_{i}}\right]\,,\qquad\mathbf{E}=-\partial_{t}\mathbf{A}\,.
\end{equation}
 With Peierls substitution, the hopping terms are modified as
\begin{equation}
t_{ij}\to t_{ij}e^{ie\phi_{ij}}\,,\qquad\phi_{ij}=A_{x}\left(\mathbf{r}_{i},t\right)\left(x_{i}-x_{j}\right)\,.
\end{equation}
 Here we assume the slowly-varying electric field is applied in $x$-direction
and the hopping is short-ranged. Then the attractive Hubbard model
under vector field is given by
\begin{equation}
H=H_{0}-e\sum_{i}j_{x}^{P}\left(\mathbf{r}_{i}\right)A_{x}\left(\mathbf{r}_{i}\right)-\frac{e^{2}}{2}\sum_{i}A_{x}^{2}\left(\mathbf{r}_{i}^{2}\right)K_{x}\left(\mathbf{r}_{i}\right)\,.
\end{equation}
The paramagnetic current and kinetic-energy densities are
\begin{align}
j_{x}^{P}\left(\mathbf{r}_{i}\right) & =i\sum_{j\sigma}t_{ij}\left(x_{i}-x_{j}\right)c_{i\sigma}^{\dagger}c_{j\sigma}\,,\\
K_{x}\left(\mathbf{r}_{i}\right) & =-\sum_{j\sigma}t_{ij}\left(x_{i}-x_{j}\right)^{2}c_{i\sigma}^{\dagger}c_{j\sigma}\,.
\end{align}
 The total current is given by 
\begin{equation}
j_{x}\left(\mathbf{r}_{i}\right)=-\frac{\delta H}{\delta A_{x}\left(\mathbf{r}_{i}\right)}=eJ_{x}^{P}\left(\mathbf{r}_{i}\right)+e^{2}A_{x}\left(\mathbf{r}_{i}\right)K_{x}\left(\mathbf{r}_{i}\right)\,.
\end{equation}
 With linear response theory, we have 
\begin{align}
J_{x}^{P}\left(\omega,\mathbf{q}\right) & =-e\Pi_{xx}^{R}\left(\omega,\mathbf{q}\right)A_{x}\left(\omega,\mathbf{q}\right)\,,\\
j_{x}\left(\omega,\mathbf{q}\right) & =e^{2}\left[-\Pi_{xx}^{R}\left(\omega,\mathbf{q}\right)+K_{x}\right]A_{x}\left(\omega,\mathbf{q}\right)\,.
\end{align}
\begin{widetext}
 The superfluid stiffness is then given by \cite{Scalapino1992}
\begin{equation}
\frac{D_{s}}{\pi e^{2}}=\Pi_{xx}^{R}\left(\omega=0,q_{x}=0,q_{y}\to0\right)-\left\langle K_{x}\right\rangle \,.
\end{equation}
It can be shown that the superfluid stiffness for the lattice model
can be evaluated via
\begin{subequations}
\begin{align}
\frac{D_{s}}{\pi e^{2}} & =-\Pi_{xx}^{R}\left(\omega=0,q_{x}=0,q_{y}\to0\right)-\left\langle K_{x}\right\rangle \,,
\\
\Pi_{xx}^{R}\left(\omega,\mathbf{q}\right) & =\frac{2}{V}\sum_{nm}\frac{f\left(E_{n}\right)-f\left(E_{m}\right)}{\omega+E_{n}-E_{m}+i\eta}\bra{n}J_{e}^{x}\left(\mathbf{q}\right)\ket{m}\bra{m}J^{x}\left(-\mathbf{q}\right)\ket{n}\,,
\\
\hat{J}_{e}^{x}\left(\mathbf{q}\right) & =i\sum_{ij}e^{-i\mathbf{q}\cdot\mathbf{r}_{i}}t_{ij}\left(x_{i}-x_{j}\right)\ket{i\uparrow}\bra{j\uparrow}\,,
\\
\hat{J}^{x}\left(\mathbf{q}\right) & =i\sum_{ij}e^{-i\mathbf{q}\cdot\mathbf{r}_{i}}t_{ij}\left(x_{i}-x_{j}\right)\ket{i\uparrow}\bra{j\uparrow}+i\sum_{ij}e^{-i\mathbf{q}\cdot\mathbf{r}_{i}}t_{ij}\left(x_{i}-x_{j}\right)\ket{i\downarrow}\bra{j\downarrow}\,,
\\
\left\langle K_{x}\right\rangle  & =-\frac{2}{V}\sum_{ij}t_{ij}\left(x_{i}-x_{j}\right)^{2}\sum_{n}\left(u_{n}^{i}\right)^{*}u_{n}^{j}f\left(E_{n}\right)=\sum_{n}\bra{n}\hat{K}_{x}f\left(E_{n}\right)\ket{n}\,,
\\
\hat{K}_{x} & =-\frac{2}{V}\sum_{ij}t_{ij}\left(x_{i}-x_{j}\right)^{2}\ket{i\uparrow}\bra{j\uparrow}\,.
\end{align}
\end{subequations}

\subsubsection{KPM implementation}

We define the two-point spectral function,
\begin{equation}
\pi\left(\omega_{1},\omega_{2},\mathbf{q}\right)=\frac{2}{V}\sum_{nm}\bra{n}\hat{J}_{e}^{x}\left(\mathbf{q}\right)\ket{m}\bra{m}\hat{J}^{x}\left(-\mathbf{q}\right)\ket{n}\delta\left(\omega_{1}-E_{n}\right)\delta\left(\omega_{2}-E_{m}\right)\,.
\end{equation}
\end{widetext}
Then the rescaled spectral function can be expressed using the kernel
polynomial method,
\begin{equation}
\tilde{\pi}\left(\omega_{1},\omega_{2}\right)=\sum_{kl=0}^{N_{\mathcal{C}}-1}\frac{\mu_{kl}h_{kl}g_{k}g_{l}T_{k}\left(\omega_{1}\right)T_{l}\left(\omega_{2}\right)}{\pi^{2}\sqrt{\left(1-\omega_{1}^{2}\right)\left(1-\omega_{2}^{2}\right)}}\,.\label{eq:Pixx(1,2)}
\end{equation}
Here $\tilde{\pi}$ stands for the function with argument rescaled
to the interval $\left[-1,1\right]$ and 
\begin{equation}
h_{kl}=\frac{2}{1+\delta_{k,0}}\frac{2}{1+\delta_{l,0}}\,,
\end{equation}
and $g_{k}$ are the kernel damping factors.
We use the Jackson kernel for the spectral function.

The moments are obtained with 
\begin{align}
    \mu_{kl} 
    & =
    \int_{-1}^{1}d\omega_{1}\int_{-1}^{1}d\omega_{2}\tilde{\pi}_{xx}\left(\omega_{1},\omega_{2}\right)T_{k}\left(\omega_{1}\right)T_{l}\left(\omega_{2}\right)
\nonumber \\
    =&\,
    \frac{2}{N}\sum_{nm}\bra{n}\hat{J}_{e}^{x}\left(\mathbf{q}\right)\ket{m}\!\bra{m}\hat{J}^{x}\left(-\mathbf{q}\right)\ket{n}T_{k}\left(\tilde{E}_{n}\right)T_{l}\left(\tilde{E}_{m}\right)
\nonumber \\
    =&\,
    \frac{2}{V}\sum_{nm}\bra{n}T_{k}\left(\tilde{H}\right)\hat{J}_{e}^{x}\left(\mathbf{q}\right)\ket{m}\!\bra{m}T_{l}\left(\tilde{H}\right)\hat{J}^{x}\left(-\mathbf{q}\right)\ket{n}
\nonumber \\
    =&\,
    \frac{2}{V}\mathsf{Tr}\left[T_{k}\left(\tilde{H}\right)\hat{J}_{e}^{x}\left(\mathbf{q}\right)T_{l}\left(\tilde{H}\right)\hat{J}^{x}\left(-\mathbf{q}\right)\right]\,.
\end{align}
 The trace can be efficiently and accurately averaged over a small
number of random vectors.

In practice, the evaluation of $\tilde{\Pi}_{xx}$ in Eq.~(\ref{eq:Pixx(1,2)})
can be efficiently implemented with the help of the discrete FFT, similar
to the one-point case. We would evaluate $\tilde{\Pi}_{xx}\left(\omega_{1},\omega_{2}\right)$
at discrete points,
\begin{align}
    \omega_{1}=\omega_{m}
    =&\,
    \cos\left[\frac{\pi\left(m+1/2\right)}{N_{p}}\right]\,,
\nonumber\\
    \omega_{2}=\omega_{n}
    =&\,
    \cos\left[\frac{\pi\left(n+1/2\right)}{N_{p}}\right]\,.
\end{align}
Here $N_{p}$ is the number of points evaluated for the integral and
a reasonable choice is $N_{p}=2N_{\mathcal{C}}$. Then we have 
\begin{align}
    \gamma_{mn}
    =&\,
    \pi^{2}\sqrt{\left(1-\omega_{m}^{2}\right)\left(1-\omega_{n}^{2}\right)}
    \,
    \tilde{\pi}\left(\omega_{m},\omega_{n}\right)
\nonumber\\
    =&\,
    \sum_{kl=0}^{N_{\mathcal{C}}-1}
    \mu_{kl} h_{kl} g_{k} g_{l}
\nonumber\\
    &\,
    \qquad
    \times
    \cos\left[\frac{\pi k\left(m+\frac{1}{2}\right)}{N_{p}}\right]
    \cos\left[\frac{\pi l\left(n+\frac{1}{2}\right)}{N_{p}}\right].\!\!
\end{align}
 The summation is actually the discrete Fourier transform of the 2D
array $\tilde{\mu}_{kl}=\mu_{kl}g_{k}g_{l}$, and it can be efficiently
evaluated 
via the 
fast-Fourier transform. 
Thus we have 
\begin{align}
\Pi_{xx}\left(\omega,\mathbf{q}\right) & =\int_{-1}^{1}d\omega_{1}\int_{-1}^{1}d\omega_{2}\frac{f\left(\omega_{1}\right)-f\left(\omega_{2}\right)}{\omega+\omega_{1}-\omega_{2}+i\eta}\tilde{\pi}\left(\omega_{1},\omega_{2}\right)\nonumber \\
 & =\frac{1}{N_{p}^{2}}\sum_{mn=0}^{N_{p}-1}\gamma_{mn}\frac{f\left(\omega_{n}\right)-f\left(\omega_{m}\right)}{\omega+\omega_{n}-\omega_{m}+i\eta}\,.
\end{align}
Here the integral is evaluated with the Gaussian quadrature. 

For the kinetic energy term, we can defined the kinetic spectral function,
\begin{equation}
k_{x}\left(\omega\right)=\sum_{n}\bra{n}\hat{K}_{x}\ket{n}\delta\left(\omega-E_{n}\right)\,.
\end{equation}
 This term can also be evaluated with the KPM in the same fashion.
At the end of the day, we would have 
\begin{equation}
\left\langle K_{x}\right\rangle =\frac{1}{N_{p}}\sum_{m=0}^{N_{p}-1}\gamma_{m}f\left(\omega_{m}\right)\,.
\end{equation}

\begin{figure}
\centering

\includegraphics[width=\columnwidth]{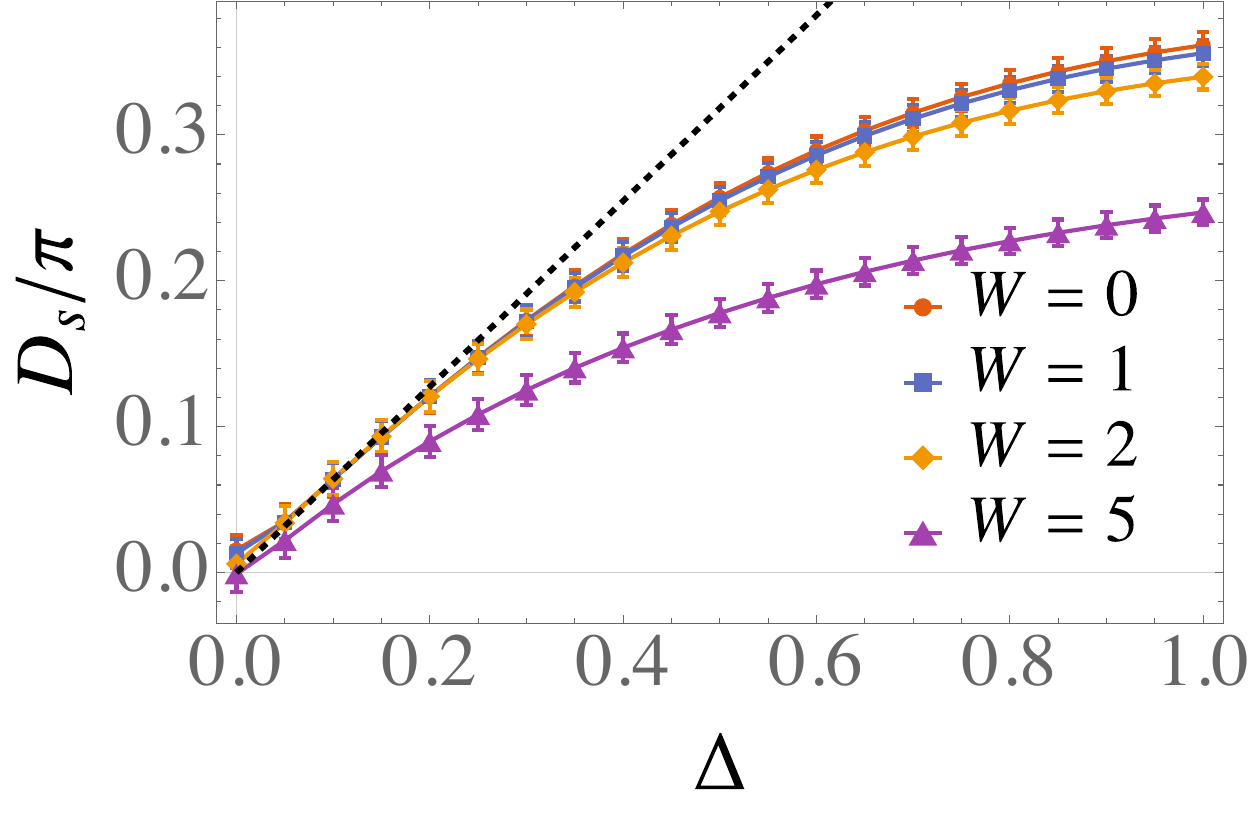}

\caption{Superfluid stiffness of 
a spin-singlet
$s$-wave superconductor on the honeycomb
lattice with uniform distributed 
on-site
disorder potential $W_{i}\in\left[-\frac{W}{2},\frac{W}{2}\right]$.
The numerical data are evaluated on a honeycomb lattice with $100\times100$
unit cells and $N_{\mathcal{C}}=1024$. The dashed line shows the
analytical prediction $\frac{D_{s}}{\pi}=\frac{2}{\pi}\Delta$ for
a clean 2D
Dirac superconductor.}
\label{fig:stiffness-honeycomb}
\end{figure}

In Fig.~\ref{fig:stiffness-honeycomb}, we show the superfluid stiffness
of a Dirac superconductor on the honeycomb lattice with onsite disorder
potential evaluated with the KPM. The superfluid stiffness $D_{s}/\pi$
follows the analytical prediction $D_{s}/\pi=2\Delta/\pi$ when the
pairing amplitude is weak in the clean system. Interestingly, the
superfluid stiffness is robust against a weak disorder potential and
is only weakened by strong disorder.

\subsubsection{Superfluid stiffness for Dirac superconductor}

Now we consider Dirac superconductor with spin-singlet, layer-singlet,
pseudospin-singlet pairing,
\begin{align}
    H
    =&\,
    \sum_{\mathbf{r}}c_{\mathbf{r}}^{\dagger}\left(-iv_{F}\boldsymbol{\sigma}\cdot\nabla\right)c_{\mathbf{r}}
\nonumber\\
    &\,
    +
    \Delta\sum_{\mathbf{r}}
    \left[ic_{\mathbf{r}\uparrow}^{\dagger}\sigma^{2}\kappa^{2}\left(c_{\mathbf{r}\downarrow}^{\dagger}\right)^{\mathsf{T}}+\textrm{H.c.}\right]\,.
\end{align}
Here $c_{\mathbf{r}}$ 
carries
pseudospin ($\sigma$),
layer ($\tau$),
and physical spin-1/2 ($s$) indices.
Introducing the Nambu spinor 
\begin{equation}
\chi=\left[\begin{array}{c}
c_{\uparrow}\\
i\sigma^{2}\kappa^{2}\left(c_{\downarrow}^{\dagger}\right)^{\mathsf{T}}
\end{array}\right]_{\mu}\,,
\end{equation}
we have 
\begin{equation}
    H
    =
    \sum_{\mathbf{r}}
    \chi^{\dagger} \, \hat{h}_{\mathsf{BdG}}\, \chi\,,
    \qquad
    \hat{h}_{\mathsf{BdG}}
    =
    \left(-iv_{F}\boldsymbol{\sigma}\cdot\nabla\right)\mu^{3}+\Delta\mu^{1}\,.
\end{equation}
Here $\mu^{i}$ is the Pauli matrix in the particle-hole sector.
The electric current operator is then given by
\begin{equation}
\mathbf{J}=\chi^{\dagger}\boldsymbol{\sigma}\chi\,.
\end{equation}
The current-current correlation function is then given by
\begin{align}
    \Pi_{xx}^{R}\left(\omega,\mathbf{q}\right)
    =&\,
    \frac{1}{V}\sum_{nm}\frac{f\left(E_{n}\right)-f\left(E_{m}\right)}{\omega+E_{n}-E_{m}+i\eta}
\nonumber\\
    &\,
    \qquad\times
    \bra{n}J_{x}\left(\mathbf{q}\right)\ket{m}\bra{m}J_{x}\left(-\mathbf{q}\right)\ket{n}\,,
\end{align}
which can be evaluated with KPM in the same fashion as the lattice
model. However, unlike the lattice model, the diamagnetic term $\left\langle K_{x}\right\rangle $
vanishes in the expression for the superfluid stiffness and the paramagnetic
current-current correlation functions diverges if no momentum cutoff
is
enforced. The superfluid stiffness should vanish when $\Delta=0$,
i.e.\ there is no Meissner effect for normal state. Therefore, we
can regularize the current-current correlation function to obtain
the physical superfluid stiffness via
\begin{align}
    \frac{D_{s}}{\pi}
    =&\,
    -\Pi_{xx}^{R}\left(\omega=0,q_{x}=0,q_{y}\to0,\Delta\neq0\right)
\nonumber\\&\,
    +\Pi_{xx}^{R}\left(\omega=0,q_{x}=0,q_{y}\to0,\Delta=0\right)\,.
\end{align}


\section{More Details on DOS and critical states \label{sec:MoreDetail}}

\subsection{System size scaling of multifractal wave function}


\begin{figure*}[tp]
  \centering
  \includegraphics[width=0.95\textwidth]{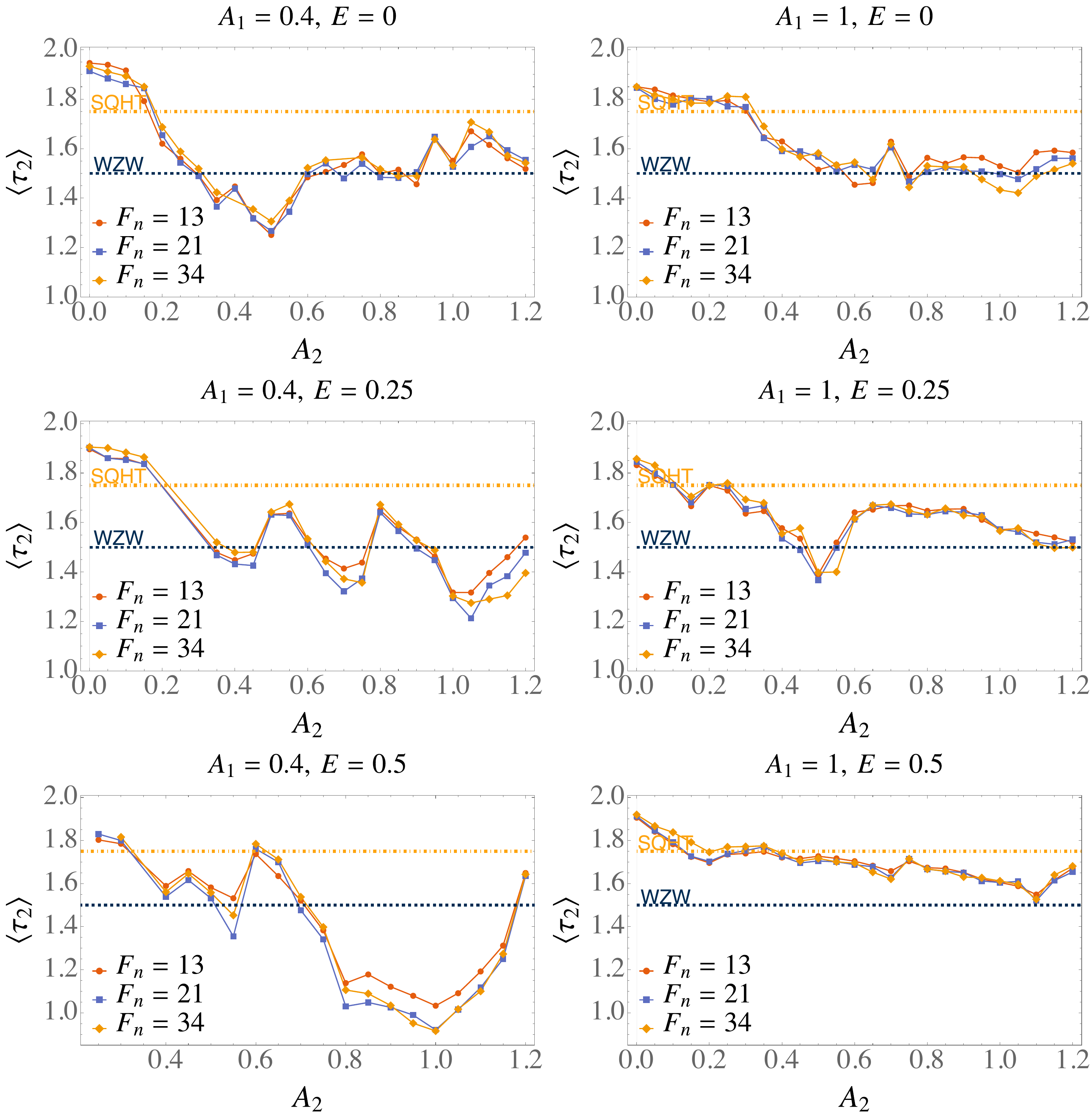}
  \caption{The multifractal dimension $\tau_2$ of wave functions at different energy with changing system size $F_n$ along the two cuts $A_1=0.4$ and $1$.
 }
  \label{fig:tau2-M}
\end{figure*}

Fig.~\ref{fig:tau2-M} shows the second multifractal dimension $\tau_2$ with different system sizes. It can be seen that $\tau_2$ converges to the same value for increasing system sizes at most points, indicating that the critical wave functions in the qBM model are NOT a finite-size effect. There are several points where the $\tau_2$ shows larger deviations for different system sizes. This occurs close to the filaments and the deviations can be attributed to the numerical errors when there are large number of states with similiar energies (filaments).


\subsection{DOS and multifractal properties in the lakes between filaments}

\begin{figure*}[tp]
  \centering
  \includegraphics[width=0.95\textwidth]{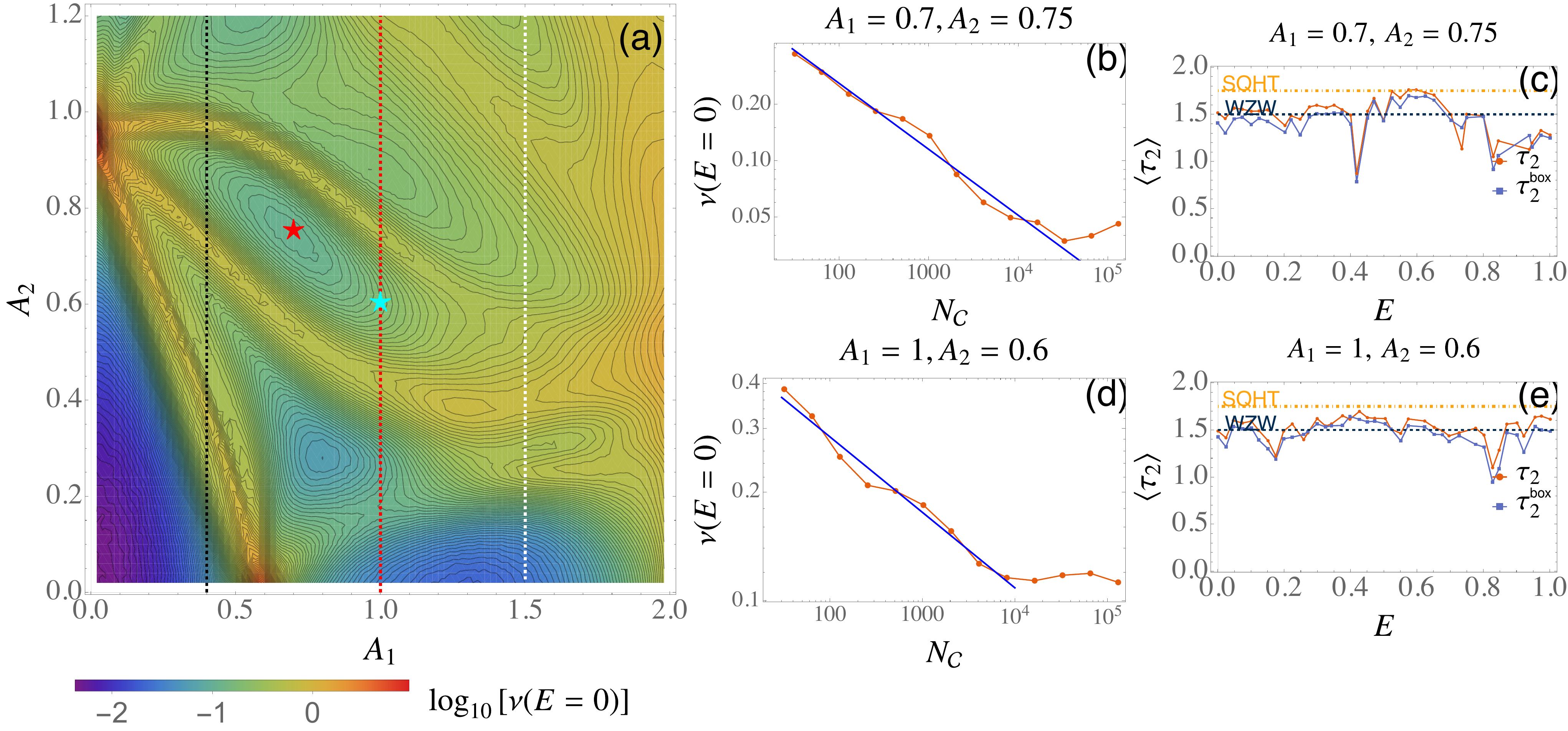}
  \caption{The zero energy DOS $\nu(E=0)$, $N_\mathcal{C}$ scaling of $\nu(E=0)$ and multifractal dimension of wave functions in the lake of DOS map. In (b) and (d), we have $F_n=233$ and $N_\Lambda = 15$ and the $N_\mathcal{C}$ dependence are fitted with $\sim N_\mathcal{C}^{-0.352}$ and $\sim N_\mathcal{C}^{-0.209}$, respectively. 
 }
  \label{fig:lake}
\end{figure*}

The density of states of a system is given by 
\begin{equation}
    \rho\left(E\right)=\frac{1}{L}\sum_{i}\delta\left(E-\varepsilon_{i}\right)\,,
\end{equation}
with 
$\{\varepsilon_{i}\}$ denoting the set of eigenenergies.
In 
the
KPM \cite{KPM2006}, the Dirac delta functions are approximated by smooth functions.
Using the Jackson kernel, 
these are Gaussians
\begin{equation}
    \delta_{\text{KPM}}^{J}\left(x\right)
    \approx
    \frac{1}{\sqrt{2\pi\sigma^{2}}}
    \exp\left[
        -\frac{\left(x-a\right)^{2}}{2\sigma^{2}}
        \right]\,.
\end{equation}
The variance
\begin{equation}
\sigma^{2}=\left(\frac{\pi}{N_{\mathcal{C}}}\right)^{2}\left[1-a^{2}+\frac{3a^{2}-2}{N_\mathcal{C}}\right]\,.
\end{equation}
Here $-1\le a\le1$ and $\sigma=\pi/N_{\mathcal{C}}$ at $a=0$,
$\sigma=\pi/N_{\mathcal{C}}^{3/2}$ at $a=\pm1$, with $N_{\mathcal{C}}$
being the order of 
the
Chebyshev expansion.

Suppose we have a system whose density of states has power-law dependence
on energy near $E=0$,
\begin{equation}
\rho\left(E\right)\propto E^{\alpha}\,.
\end{equation}
Then in the KPM with Jackson kernel, we have 
\begin{align}
    \rho_{KPM}\left(E=0\right) 
    \approx&\,
    \frac{1}{\sqrt{2\pi\sigma^{2}}}
    \int dE \,
    \rho\left(E\right)e^{-\frac{E^{2}}{2\sigma^{2}}}
\nonumber\\
    =&\,
    \frac{1}{\sqrt{2\pi\sigma^{2}}}
    \int dE \, E^{\alpha}
    e^{-\frac{E^{2}}{2\sigma^{2}}}
\nonumber\\
    =&\,
    \frac{\sigma^{\alpha+1}}{\sqrt{2\pi\sigma^{2}}}
    \int dx \, x^{\alpha}e^{-\frac{x^{2}}{2}}
    \propto\sigma^{\alpha}\sim N_{\mathcal{C}}^{-\alpha}\,.
\end{align}
The 
scaling
behavior holds when $N_\mathcal{C}$ is not too small or too large. When $N_\mathcal{C}$ is too small, the higher-order corrections are significant and spoil the simple power-law scaling. When $N_\mathcal{C}$ is large enough that $\sigma$ is at the order of level spacing of the system, the DOS 
resolves into isolated Gaussian peaks.

Fig.~\ref{fig:lake} shows the $N_\mathcal{C}$ scaling of the zero-energy DOS and multifractal dimensions of the wave functions at two points 
$A_1=0.7$, $A_2=0.75$ and $A_1=1$, $A_2=0.6$ in the lakes between the filaments.
When $N_\mathcal{C}$ is not very large, the $\nu(E=0)$ has power-law dependence on $N_\mathcal{C}$. It implies that the DOS near zero energy has power-law dependence on energy with 
$\nu(E)\sim \left|E\right|^{0.352}$ for $A_1=0.7$, $A_2=0.75$, 
and 
$\nu(E) \sim \left|E\right|^{0.209}$ for $A_1=1$, $A_2=0.6$, respectively.

\begin{figure*}[tp]
  \centering
  \includegraphics[width=0.8\textwidth]{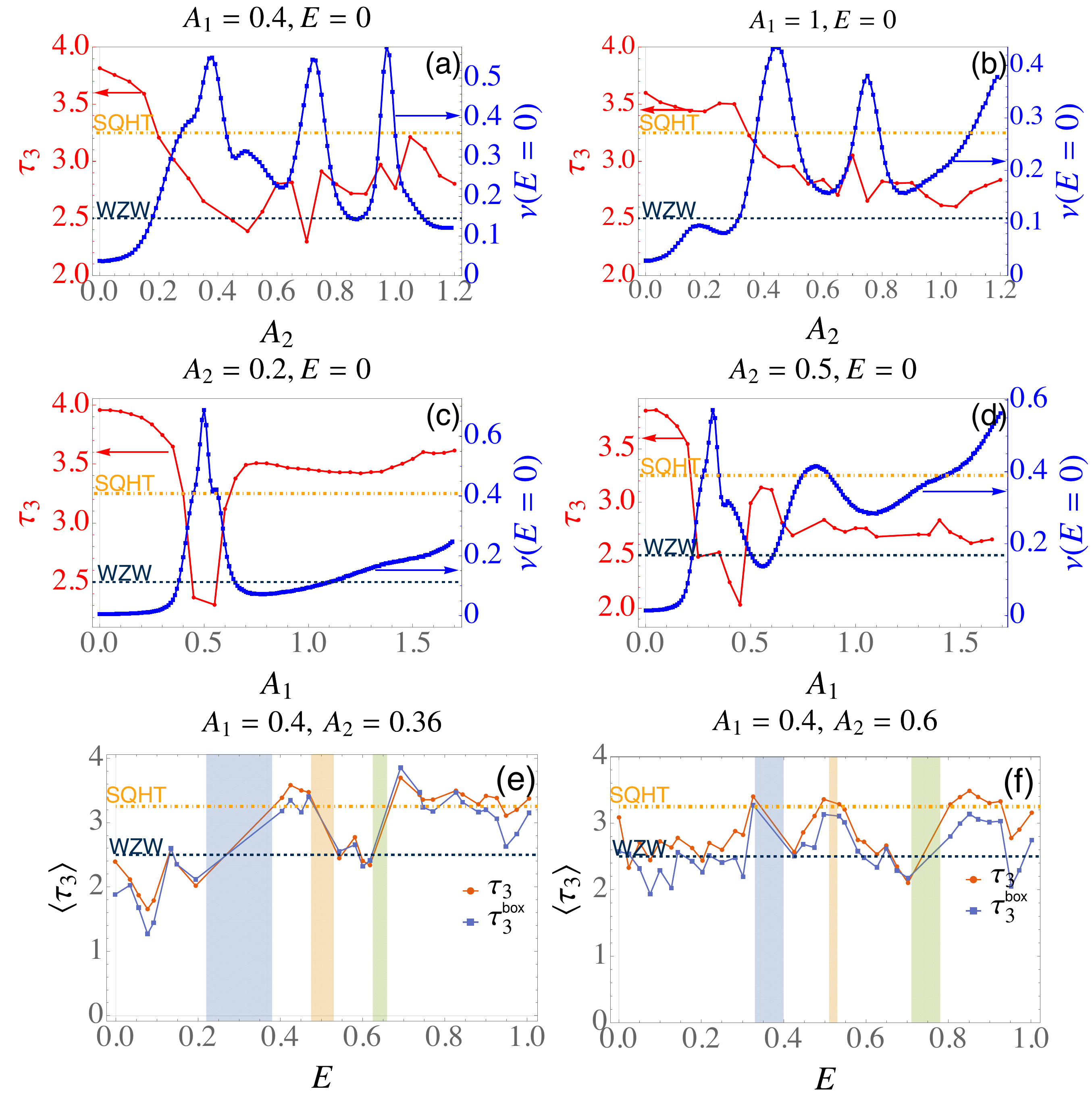}
  \caption{The third multifractal dimension $\tau_3$ averaged for states near zero energy
  (a)--(d) or finite energies (e), (f), along with the zero-energy DOS 
  $\nu(E=0)$ for the golden-ratio qBM model. 
  (a) and (b): along the vertical cuts at $A_1=0.4$ and $1$, respectively; 
  (c) and (d) along the horizontal cuts at $A_2=0.2$ and $0.5$, respectively; 
  (e) and (f) $\tau_3$ versus energy for $A_1=0.4$ and $A_2=0.36$ (at a filament), 
  $A_2=0.6$ (in the lake between filaments). 
  The navy blue dashed and orange dot-dashed lines indicate $\tau_3=5/2$ for WZW states and $13/4$ for SQHT states, respectively \cite{Ghorashi18}.  
 }
  \label{fig:tau3}
\end{figure*}

\subsection{The third multifractal dimension $\tau_3$}

In Figs.~\ref{fig:tau2-A2} and \ref{fig:tau2-wf}, we show the second multifractal dimension $\tau_2$ for the qBM model. Fig.~\ref{fig:tau3} shows the third multifractal dimension $\tau_3$.


\section{Silver Ratio QBM Model \label{sec:SilverRatio}}

\begin{figure*}[tp]
  \centering
  \includegraphics[width=0.9\textwidth]{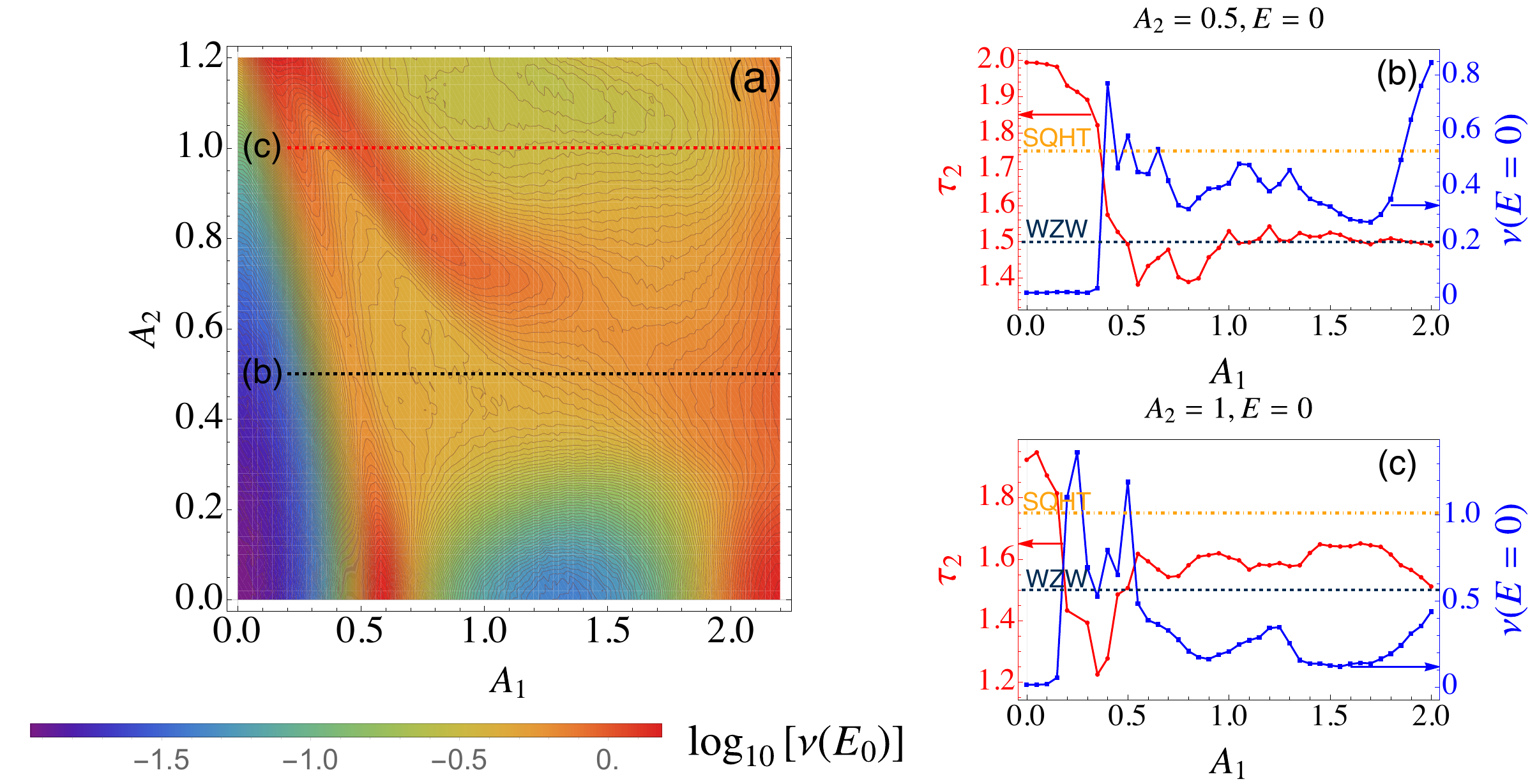}
  \caption{Zero-energy DOS and the second multifractal dimension for the silver-ratio qBM model with $\mathbf{q}_i = \beta \, \mathbf{q}_i'$, with $\beta$ 
  a commensurate approximant to
  the inverse silver ratio $1/\left(\sqrt{2}+1\right)$.  
 }
  \label{fig:nuE0_SR}
\end{figure*}

The critical filaments and multifractal properties of wave functions are universal for 
the
quasiperiodic Bistritzer-MacDonald model. In 
Secs.~\ref{sec:MainResults} and \ref{sec:MoreDetail},
we focus on the case 
$\mathbf{q}_i=\beta \, \mathbf{q}'_i$ with $\beta$ the inverse golden ratio. 
In this section, 
we consider instead $\beta = 1/(\sqrt{2} + 1)$, which is the inverse silver ratio.
The 
silver ratio 
can be approximately by $P_{n+1}/P_{n}$ with $P_{n} = 1, 2, 5, 12, 29, 70, \dots$ the Pell numbers.
Fig.~\ref{fig:nuE0_SR} shows the zero-energy DOS map, and the second multifractal dimension $\tau_2$ along two horizontal cuts $A_2=0.5$ and $A_2=1$ for the silver-ratio qBM model. There are filaments with large-zero energy DOS connecting the magic angle points of $A_1$ and $A_2$. Between the filaments, the 
nonzero
zero-energy DOS shows that the system is no longer semimetallic. At the filaments and in the lakes between the filaments, the wave functions near the zero energy become multifractal. The $\tau_2$ is close to that of WZW state ($\tau_2=3/2$) for zero-energy wave functions in the lake along the cut $A_1=0.5$. In the lake along the $A_2=1$ cut, $\tau_2$ is between that of WZW and SQHT states ($\tau_2=7/4$). At the filaments, $\tau_2$ is smaller than $3/2$, indicating that these states are strongly multifractal. The results are 
very
similar to the case of golden-ratio qBM model, indicating that our 
conclusions 
are general.


\section{More details on the superconductivity \label{sec:SCDetails}}

\subsection{Numerical results}

We consider superconductivity 
in the
qBM model with a local pairing interaction,
\begin{equation}
    H 
    = 
    \int d\mathbf{r} \,
    \psi^\dagger(\mathbf{r})
    \,
    h(\mathbf{r})
    \,
    \psi(\mathbf{r})
    - 
    U 
    \int d\mathbf{r} \, 
    \mathcal{B}^\dagger
    \,
    \mathcal{B}\,.
    \label{eq:H_U}
\end{equation}
Here $h(\mathbf{r})$ is the qBM 
Hamiltonian given by
Eq.~(\ref{eq:hqBM}), 
$\psi\rightarrow\psi_{\sigma,\tau,s}$ carries indices in sublattice ($\sigma$), layer ($\tau$), and physical spin-1/2 ($s$) spaces,
$U$ gives the strength of the local attractive interaction, and $\mathcal{B}$ is the bilinear for 
pseudospin-singlet, layer-singlet, and spin-singlet
$s$-wave pairing
\begin{equation}
    \mathcal{B} = \psi^{\mathsf{T}} \sigma^2 \tau^2 s^2 \psi.
    \label{eq:B-pairing}
\end{equation}
The local interaction is well-known to be RG irrelevant for 2+1-D Dirac fermions and superconductivity only arises for
an
interaction strength comparable to the ultraviolet energy cutoff,
see Eq.~(\ref{eq:Uc}) below.
In the presence of the usual Bistritzer-MacDonald (monochromatic) moir\'{e} potential, the low-energy spectrum remains linear and strong interactions are required for superconductivity, except in the close vicinity of the magic angle where a nearly flat band results in a much smaller critical interaction strength. 

The attractive interaction can be decoupled with the mean-field ansatz,
\begin{equation}
       \Delta(\mathbf{r}) = -U \left< \mathcal{B}(\mathbf{r})\right>.
        \label{eq:MF}
\end{equation}
Because we work with the quasiperiodic variant of the single-valley BM model, the $s$-wave pairing considered here would correspond to an FFLO state relative to the microscopic honeycomb lattices. We expect that the physics of pairing enhancement due to Chalker scaling of quantum-critical wave functions \cite{Feigelman07,Feigelman10}
and (approximate) topology-protected criticality will hold for intervalley pairing as well, but we leave this as a study for future work.

The mean-field Eqs.~(\ref{eq:hBdG}) and (\ref{eq:MF}) 
are
then readily solved self-consistently with the KPM based method. The KPM-based method allows us to deal with system sizes
inaccessible with exact diagonalization. 
In Fig.~\ref{fig:SC},
the data are obtained with the commensurate fraction denominator $F_n=8$ and $N_{\Lambda}=11$, and the dimension
of the Hamiltonian matrix is $61952$, which is very hard to 
diagonalize exactly
even with the help of modern GPUs. 

\begin{figure*}[tp]
  \centering
  \includegraphics[width=0.9\textwidth]{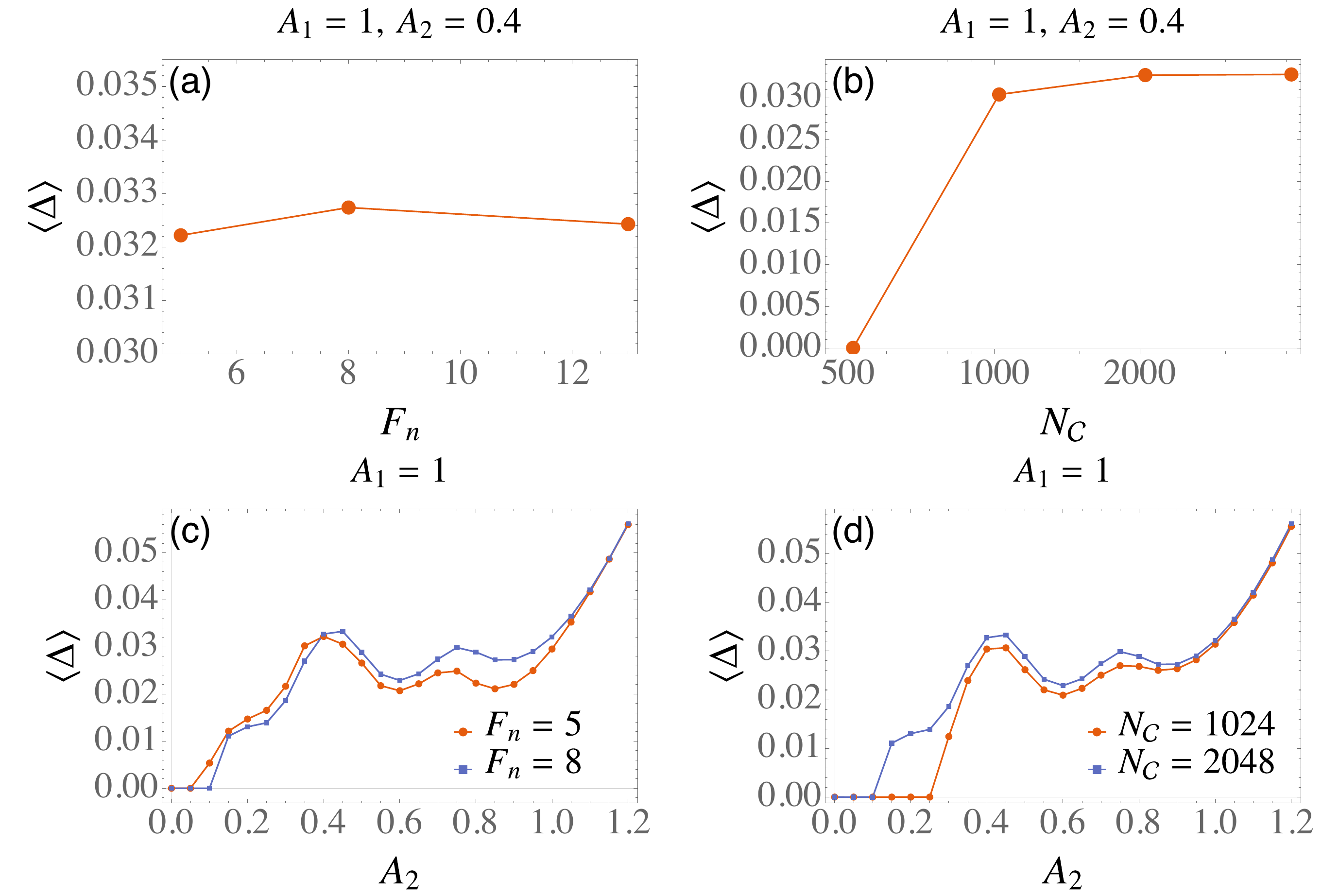}
  \caption{The average pairing amplitude for qBM-BCS model with different system size and maximal degree of kernel polynomials $N_\mathcal{C}$.
 }
  \label{fig:sc-M-Nc}
\end{figure*}

Fig.~\ref{fig:sc-M-Nc} shows the dependence of spatially averaged pairing amplitude $\left<\Delta\right>$ on the system size and the degree of the kernel polynomials. We show the data with three system sizes 
$F_n=5$, $8$, and $F_n=13$ with $N_\Lambda=11$, which is solved on a triangular lattice with $55\times 55$, $88\times 88$ and $143\times 143$ sites, respectively. The $\left<\Delta\right>$ only slightly changes with $F_n$. It indicates that the system size employed is large enough and the results apply for systems in thermodynamic limit. Fig.~\ref{fig:sc-M-Nc}(b) show the $N_\mathcal{C}$ dependence of $\left<\Delta\right>$ and the results converges when $N_\mathcal{C}$ is larger than $10^{3}$. In Fig.~\ref{fig:SC}, we use $F_n=8$, $N_\Lambda=11$ and $N_\mathcal{C}=2048$.

\subsection{Pairing of exact eigenstates approximation \label{sec:PoEE}}

\begin{widetext}
Consider Dirac fermions with local attractive interactions and a spatially inhomogeneous potential,
\begin{equation}\label{HPoEE}
    H =
    \sum_{\mathbf{r}}
    c_{\mathbf{r}}^{\dagger}
    \,
    h(\vex{r})
    \,
    c_{\mathbf{r}}
    -
    U\sum_{\mathbf{r}}
    \left[
    c_{\mathbf{r}\uparrow}^{\dagger}
    \left(
    -i\sigma^{2}\tau^{2}
    \right)
    \left(
    c_{\mathbf{r}\downarrow}^{\dagger}
    \right)^{\mathsf{T}}
    \right]
    \left[
    \left(
    c_{\mathbf{r}\downarrow}
    \right)^{\mathsf{T}}
    i\sigma^{2}\tau^{2}c_{\mathbf{r}\uparrow}
    \right]
    \,.
\end{equation}
Here $c_{\mathbf{r}}$ carries pseudospin ($\sigma$), layer ($\tau$)
and spin ($s$) indices, 
and the one-body Dirac Hamiltonian $h(\vex{r})$ incorporates
random or structured matrix potentials in some symmetry class, e.g.\ 
the class-CI chiral qBM model in Eq.~(\ref{eq:hqBM}).
As in that case, we assume spin SU(2) and time-reversal symmetries in the 
normal state.
Suppose the non-interacting part is solved
by 
\begin{equation}
    h
    \ket{\alpha}
    =
    \varepsilon_{\alpha}
    \ket{\alpha}\,.
\end{equation}
Then we have 
\begin{equation}
c_{\mathbf{r} \sigma \tau s}
    =\sum_{\alpha}\braket{\mathbf{r\sigma}\tau|\alpha}a_{\alpha s}
    =\sum_{\alpha}\psi_{\alpha}\left(\mathbf{r}\sigma\tau\right)a_{\alpha s}\,.
\end{equation}
Expressing the Eq.~(\ref{HPoEE}) in this basis gives
\begin{equation}\label{HPoEE2}
    H
    =\sum_{\alpha,s}\varepsilon_{\alpha}a_{\alpha s}^{\dagger}a_{\alpha s}
    -U\sum_{\mathbf{r}}\sum_{\alpha\beta\gamma\delta}\left[\psi_{\alpha}^{\dagger}\left(\mathbf{r}\right)\left(-i\sigma^{2}\tau^{2}\right)\left(\psi_{\beta}^{^{\dagger}}\left(\mathbf{r}\right)\right)^{\mathsf{T}}\psi_{\gamma}^{\mathsf{T}}\left(\mathbf{r}\right)\left(i\sigma^{2}\tau^{2}\right)\psi_{\delta}\left(\mathbf{r}\right)\right]
    a_{\alpha\uparrow}^{\dagger}a_{\beta\downarrow}^{\dagger}a_{\gamma\downarrow}a_{\delta\uparrow}\,.
\end{equation}
The Cooper pairs are formed by 
pairing the
time-reversal 
counterparts 
of spin-up and -down electrons. The time-reversed conjugate of $\ket{\alpha}$
is given by
\begin{equation}
    \ket{\bar{\alpha}}=-i\sigma^{2}\tau^{2}\ket{\alpha^{*}}\,.
\end{equation}
We neglect interactions in Eq.~(\ref{HPoEE2}) except those responsible for creating such pairs. Thus we obtain
the reduced BCS Hamiltonian 
\begin{align}
    H_{\mathsf{red}} 
    & =
    \sum_{\alpha,s}
    \varepsilon_{\alpha}
    a_{\alpha s}^{\dagger}a_{\alpha s}
    -
    U\sum_{\alpha\beta}
    \left[
        \sum_{\mathbf{r}}
        \psi_{\alpha}^{*}\left(\mathbf{r}\right)
        \left(-i\sigma^{2}\tau^{2}\right)
        \psi_{\bar{\alpha}}^{*} \left(\mathbf{r}\right) 
        \psi_{\bar{\beta}}\left(\mathbf{r}\right)
        \left(i\sigma^{2}\tau^{2}\right)
        \psi_{\beta}\left(\mathbf{r}\right)
    \right]
    a_{\alpha\uparrow}^{\dagger} a_{\bar{\alpha}\downarrow}^{\dagger}a_{\bar{\beta}\downarrow}a_{\beta\uparrow}
    \nonumber \\
     & =
     \sum_{\alpha,s}
     \varepsilon_{\alpha}
     a_{\alpha s}^{\dagger}a_{\alpha s}
     -
     \sum_{\alpha\beta}M_{\alpha\beta} \,
     a_{\alpha\uparrow}^{\dagger}a_{\bar{\alpha}\downarrow}^{\dagger}a_{\bar{\beta}\downarrow}a_{\beta\uparrow}\,.
\end{align}
\end{widetext}
Here
\begin{equation}
M_{\alpha\beta}=U\sum_{\mathbf{r}}\left|\psi_{\alpha}\left(\mathbf{r}\right)\right|^{2}\left|\psi_{\beta}\left(\mathbf{r}\right)\right|^{2}\,,
\end{equation}
where we have used
\begin{equation}
\psi_{\alpha}=-i\sigma^{2}\tau^{2}\psi_{\alpha}^{*}\,,\qquad\left|\psi_{\alpha}\left(\mathbf{r}\right)\right|^{2}\equiv\sum_{\sigma\tau}\left|\psi_{\alpha}\left(\mathbf{r}\sigma\tau\right)\right|^{2}\,.
\end{equation}
Then we have the mean-field gap equation,
\begin{equation}
\Delta_{\alpha}=\sum_{\lambda}M_{\alpha\lambda}\tanh\left(\frac{\beta\sqrt{\varepsilon_{\lambda}^{2}+\Delta_{\lambda}^{2}}}{2}\right)\frac{\Delta_{\lambda}}{2\sqrt{\varepsilon_{\lambda}^{2}+\Delta_{\lambda}^{2}}}\,.
\end{equation}

In the \emph{clean limit}, the wave functions are plane waves and the spatial overlap of wave functions at different energies is the same,
\begin{equation}
    M_{\alpha\beta} 
    \approx
    U
    \left( \frac{a}{L}\right)^d\,.
\end{equation}
Here $a$ is the lattice constant and $L$ is the system size.
The mean field gap equation is reduced 
to
\begin{equation}
    \Delta_{\alpha}
    =
    U\sum_{\lambda}\left(\frac{a}{L}\right)^{d}
    \tanh\left(\frac{\beta\sqrt{\varepsilon_{\lambda}^{2}+\Delta_{\lambda}^{2}}}{2}\right)\frac{\Delta_{\lambda}}{2\sqrt{\varepsilon_{\lambda}^{2}+\Delta_{\lambda}^{2}}}\,.
\end{equation}
Assuming that
the pairing amplitude is energy independent, $\Delta\left(\varepsilon_{\alpha}\right)=\Delta$, we have
\begin{equation}
    1\approx Ua^{d}
    \int_{-\Lambda}^{\Lambda}
    d\varepsilon
    \, \nu\left(\varepsilon\right)
    \tanh\left(\frac{\beta\sqrt{\varepsilon^{2}+\Delta^{2}}}{2}\right)
    \frac{1}{2\sqrt{\varepsilon^{2}+\Delta^{2}}}\,.
\end{equation}
For Dirac fermions in 2D, the DOS is given by
\begin{equation}
    \nu(\varepsilon)
    = 
    \frac{1}{\pi v_F^2} \left|\varepsilon\right|\,.
\end{equation}
Thus, we have the zero-temperature gap equation
\begin{align}
    \frac{2}{U}
    =&\,
    a^{2}\int_{-\Lambda}^{\Lambda}d\varepsilon\frac{\left|\varepsilon\right|}{\pi v_{F}^{2}}\frac{1}{\sqrt{\varepsilon^{2}+\Delta^{2}}}
\nonumber\\
    =&\,
    \frac{a^{2}}{\pi v_{F}^{2}}
    \left(
    \sqrt{\Lambda^{2}+\Delta^{2}}
    -\Delta
    \right)\,.
\end{align}
Here $\Lambda$ is the energy cutoff of the Dirac fermions. The equation has a solution for nonzero $\Delta$ only when the interaction strength is above the critical value
\begin{equation}
    U_{c}=\frac{2\pi v_{F}^{2}}{\Lambda a^{2}} 
    = 
    \frac{2\Lambda}{\pi}\,,
    \label{eq:Uc}
\end{equation}
where we have assumed that $\Lambda = v_F ( \pi / a)$.
Thus we obtain the well-known result that superconductivity only arises above the critical interaction strength for 
clean 2D
Dirac fermions \cite{Uchoa2005, Uchoa2007, Kopnin08, Rahul2013, IDP2014}. 

For \emph{critical states}, the spatial overlap of 
the
wave functions 
entering the matrix element
is 
determined by
Chalker scaling \cite{Foster2014, Chou2014, Karcher21}, and we have
\begin{equation}
    M_{\alpha\beta} 
    \sim 
    U 
    \mathfrak{b}
    \left|\varepsilon_\alpha - \varepsilon_\beta\right|^{(\tau_2-d)/z}\,.
\end{equation}
Here $z$ is the dynamical critical exponent of 
Dirac fermions with random or structural spatial inhomogeneity.
The DOS of critical Dirac fermions near $\varepsilon = 0$ is modified by 
the disorder or quasiperiodic potential,
\begin{equation}
    \nu(\varepsilon) \approx \mathfrak{c} \left|\varepsilon\right|^{(d-z)/z} \,.
\end{equation}
Thus we have the gap equation at $T=0$,
\begin{align}
    1 =&\, 
    U 
    \int_{-\Lambda}^{\Lambda} 
    d\varepsilon
    \,
    \nu(\varepsilon)
    \,
    M_{\alpha\beta}\tanh\left(\frac{\beta\sqrt{\varepsilon^{2}+\Delta^{2}}}{2}\right)
    \frac{1}{2\sqrt{\varepsilon^2 + \Delta^2}} 
\nonumber\\
    \approx&\, 
    U\mathfrak{a} 
    \int_{-\Lambda}^{\Lambda} 
    d\varepsilon
    \left|\varepsilon\right|^{\frac{\tau_2-z}{z}}
    \tanh\left(\frac{\beta\sqrt{\varepsilon^{2}+\Delta^{2}}}{2}\right)
    \frac{1}{2\sqrt{\varepsilon^2 + \Delta^2}}. 
\end{align}
Here $\mathfrak{a} = \mathfrak{b}\mathfrak{c}$. 
For 2D disordered Dirac fermions in the chiral/WZNW version of class CI with winding number $2$, we have 
\cite{Foster2014}
\begin{equation}
    z = \frac{7}{4}
\end{equation}
If we further assume that, due to SWQC \cite{Ghorashi18,Karcher21}, the appropriate
wave-function fractal dimension is that of the class-C SQHT
\begin{align}
    d_2 \approx 7/4,
\end{align}
then 
we have
\begin{equation}
    1 \approx  U\mathfrak{a} 
    \int_{-\Lambda}^{\Lambda} 
    d\varepsilon
    \tanh\left(\frac{\beta\sqrt{\varepsilon^{2}+\Delta^{2}}}{2}\right)
    \frac{1}{2\sqrt{\varepsilon^2 + \Delta^2}}. 
\end{equation}
This is exactly the gap equation for a BCS superconductor 
of electrons with a Fermi surface
if $\mathfrak{a}$ is replaced by the 
constant DOS at the Fermi level.
At zero temperature, we have 
\begin{equation}
    \Delta  
    = 
    \frac{\Lambda}{\sinh\left(\frac{1}{U\mathfrak{a}}\right)} 
    \approx 
    2 \Lambda \, e^{-1/(U\mathfrak{a})}.
    \label{eq:Delta-U}
\end{equation}
By contrast, if we instead employ the fractal dimension for low-energy WZNW wave functions $d_2 = 3/2$, we would get the non-BCS result if we neglect the UV cutoff \cite{Feigelman07,Feigelman10} 
\begin{align}
    \Delta \propto U^{1/7}.
\end{align}
In both scenarios, the finite-threshold strength in the clean limit $U_c$ 
[Eq.~(\ref{eq:Uc})] is lowered to zero for quantum-critical wave functions,
\emph{despite} the vanishing density-of-states for class CI $\nu(\e) \approx |\e|^{1/7}.$

\begin{figure*}[tp]
  \centering
  \includegraphics[width=0.9\textwidth]{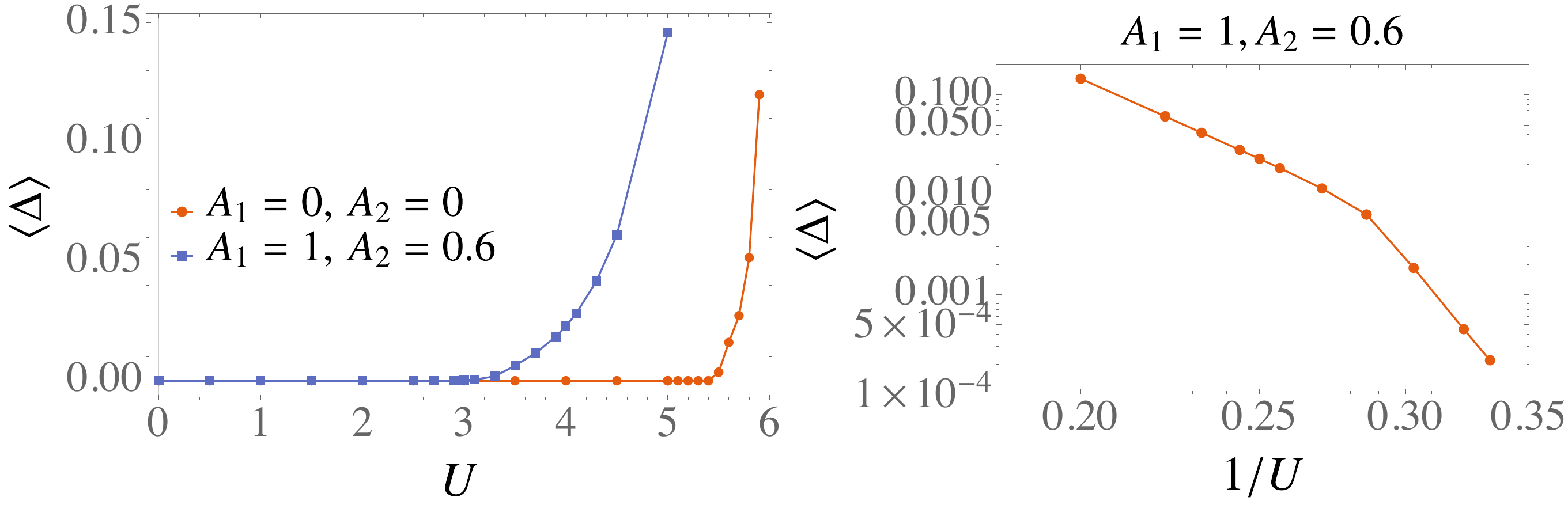}
  \caption{The dependence of the pairing amplitude $\Delta$ on the interaction strength $U$ for Dirac fermions in both the clean case and in the chiral qBM model. 
 }
  \label{fig:Delta-U}
\end{figure*}

The critical exponents in the qBM model deviate from the case of dirty Dirac fermions. But as is shown in Fig.~\ref{fig:lake}, the DOS in the lake shows power-law dependence on energy. 
In Fig.~\ref{fig:Delta-U}, we show the dependence of the pairing amplitude on the interaction strength for Dirac fermions in the
clean limit and in the critical lake of 
the chiral
qBM model. 
In 
the
clean limit, $\Delta$ becomes nonzero when $U$ is above the critical interaction strength. In the lake with $A_1=1$ and $A_2=0.6$, the $\Delta$ vs.\ $1/U$ curve is approximately linear for weak interactions in the log-linear plot, which is consistent with the prediction in Eq.~\eqref{eq:Delta-U}.

\begin{acknowledgments}
We thank S.~Gopalakrishnan and E.~K\"onig for helpful discussions.
We further thank Aaron Dunbrack for pointing out the physical system that realizes our model. This work was supported by the Welch Foundation Grant No.~C-1809 (X.~Z. and M.~S.~F.) and by the National Science Foundation under Grant No.~DMR-2238895 (J.~H.~W.).
This work was performed in part at Aspen Center for Physics, which is supported by NSF Grant No.~PHY-2210452. This work was supported in part by the Big-Data Private-Cloud Research Cyberinfrastructure MRI-award funded by NSF under grant CNS-1338099 and by Rice University's Center for Research Computing (CRC). Portions of this research were conducted with high performance computing resources provided by Louisiana State University (http://www.hpc.lsu.edu).
\end{acknowledgments}

\end{document}